\documentstyle[aps,preprint,epsfig]{revtex}
\tighten

\begin{document}

\def\mev{\hbox{\ MeV}}
 
\draft



\title{Dynamics of  quantum systems}

\author{I.~Rotter}
\address{
Max-Planck-Institut f\"ur Physik komplexer Systeme,
D-01187 Dresden, Germany}

\date{\today}

\maketitle

\vspace*{.5cm}

\begin{abstract}
A relation between the eigenvalues of an effective Hamilton operator and the
poles of the $S$ matrix is derived which holds for isolated as well as for
overlapping resonance states. The system may be a many-particle 
quantum system with
two-body forces between the constituents or it may be a quantum billiard
without any two-body forces.
Avoided  crossings of discrete states as well as of  resonance states are 
traced back to the existence of branch points in the complex plane.
Under certain conditions, these branch points appear as double poles of the
$S$ matrix. They influence the dynamics of open as well as of closed 
quantum systems. The dynamics of the two-level system is studied in detail
analytically as well as numerically.

\end{abstract}

\vspace{.5cm}

\section{Introduction}

Recently, the generic properties of many-body
quantum systems are studied with a renewed
interest. Mostly, the level distributions are compared  with those
following from random matrix ensembles.  
The generic properties are, as a rule,
well expressed in the center of the spectra where the level density is high. 
As an example, the statistical properties of the shell-model states of nuclei 
around $^{24}$Mg  are
studied a few years ago \cite{zelev} by using two-body forces which are 
obtained  by fitting the low-lying states of  different nuclei
of the $2s-1d$ shell. In the center of the spectra,
the generic properties are well expressed  in spite 
of the two-body character of the forces used in the calculations.

Another result  is obtained recently  in performing shell-model
calculations for the same systems with random two-body forces.
In spite of the random character of the forces,
the regular properties of the low-lying states 
are described quite well \cite{bertsch} in these calculations.
Further studies \cite{frank,kaplan,zel} proved the relevance of the 
results obtained and could explain in detail 
even the regular properties at the border of the spectra
obtained from random two-body forces \cite{drozdz}.
The spectral properties of the two-body random ensemble studied 30 years 
ago are reanalyzed \cite{flores}.

The spectra of microwave cavities are not determined by  two-body 
forces. Nevertheless, the calculated spectra are similar to those  from 
nuclear reactions \cite{ropepise}. They show deviations from the 
spectra obtained from random matrix theory as well as
similarities with them. Avoided level crossings play an important role.
The theoretical results obtained  are  confirmed
by experimental studies \cite{perostba}.

The effect of avoided level crossing (Landau-Zener effect)
is known and studied theoretically as well as experimentally for many years. 
It is a quite general  property of the discrete states
of a quantum system whose energies  will never cross
when there is  a certain non-vanishing interaction between them.
Instead, they  avoid crossing in energy and
their wave functions are exchanged when
traced as a function of a certain tuning parameter.
The avoided level crossings are related to the existence of 
exceptional points \cite{hemuro}. The relation between exceptional
points and the well-known diabolical points is not investigated
up to now. The relation of the latter ones
to geometrical phases is studied experimentally
\cite{dubb} as well as theoretically \cite{geom1,geom2} 
in a microwave resonator by deforming it cyclically.
The results show non-trivial phase changes when  degeneracies 
appear near to one another \cite{geom2}.
The influence of level crossings in the complex plane onto the
spectra of atoms is studied in \cite{solov1}. In this case, 
the crossings are called hidden crossings \cite{solov2}.
The relation between avoided level crossings and double poles of the
$S$ matrix is traced   in
laser-induced continuum structures in atoms \cite{marost1,marost2}.    

Usually, it is assumed that avoided level crossings do not
introduce any  correlations between the wave functions of the states
as long as the system parameter is different from the critical one 
at which the two states avoided  crossing. 
Counter-examples have been found, however, in recent numerical studies of the 
spectra of microwave cavities \cite{ropepise,reichl}. 
These results coincide with the idea \cite{haake} that avoided crossings are a 
mechanism of generating the random matrix like properties in
spectra of quantum systems.

It is the aim of the present paper to study in detail the dynamics of
quantum systems which is caused by avoided level crossings.
They are traced back to the existence of  branch points in the complex plane
where  the interaction  between the two states is maximum. 
The wave functions are bi-orthogonal in the whole function space
without any exception.

The paper is organized as follows. In section 2, the
relation between the eigenvalues of an effective Hamilton operator 
and the poles of the $S$ matrix is derived. This relation holds also
in the region of overlapping resonances. It is used 
in section 3, where the mathematical properties of branch points in the complex
plane and their relation to avoided level crossings
and double poles of the $S$ matrix are sketched
by means of a two-level model.  In section 4, numerical results 
for states at a double pole of the $S$ matrix as well as for 
discrete and resonance states with avoided  crossing
are given. The influence of  branch points in the complex plane
onto  the dynamics of  quantum systems is traced and shown to be large.  
The results are discussed and summarized in the last section.

\section{Hamilton operator and $S$ matrix}

\subsection{The wave function in the  space with discrete and
scattering states}

The solutions of the Schr\"odinger equation
\begin{eqnarray}
(H - E) \Psi_E^c = 0
\label{eq:1}
\end{eqnarray}
in the whole function space contain contributions from discrete as well as
from scattering states. The discrete states are embedded in the continuum of
scattering states and can decay. In the following, we will represent 
the solutions $\Psi_E^c$ 
by means of  the wave functions   of the discrete 
and  scattering states.

Following  \cite{baroho,ro91}, we define  two sets of wave
functions by
solving first the Schr\"odinger equation 
\begin{eqnarray}
 (H^{\rm cl} - E_R^{\rm cl})\; \Phi_R^{\rm cl} = 0
\label{eq:hamsm}
\end{eqnarray}
for the discrete  states  of the closed system
and secondly the Schr\"odinger equation 
\begin{eqnarray}  
\sum_{c'} (H^{cc'} - E) \; \xi_E^{c'(+)} = 0 
\label{eq:scatt1}
\end{eqnarray}
for the scattering states of the environment. 
Here, $H^{\rm cl}$ is the Hamilton operator for the closed system with
discrete states and $H^{cc'}$ is that for the scattering on the 
potential by which the discrete states are defined, $E$ is the energy of the
system and the channels are denoted by $c$.

By means of the two function sets
obtained, two projection operators can be defined, 
\begin{eqnarray}
Q  =  \sum_{R=1}^N \;  | \Phi_R^{\rm cl} \rangle
      \langle \Phi_R^{\rm cl} | \quad  \qquad
P  =  \sum_{c=1}^\Lambda \int_{\epsilon_c}^\infty \; dE \;
   \; | \xi_E^{c(+)} \rangle
  \langle \xi_E^{c(+)} |  
\label{eq:pqop} 
\end{eqnarray}
with    $Q \cdot \xi_E^{c(+)} = 0 \, ;  \; 
 P \cdot \Phi_R^{\rm cl} = 0 $.  
We identify $ H^{\rm cl} $  with  $ QHQ \equiv H_{QQ}$ 
and $H^{cc'}$ with $ PHP \equiv H_{PP}$. 
From (\ref{eq:1}), it follows
\begin{eqnarray}
(H_{PP} - E) \cdot P \Psi_E^{c(+)} = - H_{PQ} \cdot Q \Psi_E^{c(+)} \; ; \quad
(H_{QQ} - E) \cdot Q \Psi_E^{c(+)} = - H_{QP} \cdot P \Psi_E^{c(+)} 
\label{eq:11} 
\end{eqnarray}
and 
\begin{eqnarray}
P \Psi_E^{c(+)} = \xi_E^{c(+)} + G_P^{(+)} H_{PQ} \cdot Q \Psi_E^{c(+)} \; ;
 \quad
Q \Psi_E^{c(+)} =  (E-H^{\rm eff})^{-1} \cdot  H_{QP} \cdot  \xi_E^{c(+)} 
\label{eq:12} 
\end{eqnarray}
where $H_{PQ} \equiv PHQ$ and $H_{QP} \equiv QHP$. Further,
\begin{eqnarray}
G_P^{(+)} = P (E^{(+)} - H_{PP})^{-1} P
\label{eq:13} 
\end{eqnarray}
is the Green function in the $P$ subspace and
\begin{eqnarray}
H^{\rm eff} = H_{QQ} + H_{QP} G_P^{(+)} H_{PQ}
\label{eq:heff} 
\end{eqnarray}
is an effective Hamiltonian in the function space of discrete states.

Assuming
  $ Q + P = 1$, 
it follows from Eq. (\ref{eq:12})
\begin{eqnarray}
\Psi_E^c = (P+Q)\, \Psi_E^c =
\xi_E^c + (1+G_P H_{PQ}) \cdot Q \Psi_E^c
\label{eq:14} 
\end{eqnarray}
where the $^{(+)}$ on the wave functions are  omitted for convenience.
Using the ansatz 
\begin{eqnarray}
Q\Psi_E^c = \sum_R B_R \Phi_R^{\rm cl}
\label{eq:15} 
\end{eqnarray}
and Eq. (\ref{eq:12}), one gets
\begin{eqnarray}
B_R= \sum_{R'} 
\langle  \Phi_R^{\rm cl} | \frac{1}
{E-H^{\rm eff}} | \Phi_{R'}^{\rm cl} \rangle
\langle   \Phi_{R'}^{\rm cl} | H_{QP} | \xi_E^c \rangle
\label{eq:14a} 
\end{eqnarray}
and
\begin{eqnarray}
\Psi_E^c 
& = &  \xi_E^c + \sum_{RR'} (\Phi_R^{\rm cl} + \omega_R)
\langle  \Phi_R^{\rm cl} | \frac{1}
{E-H^{\rm eff}} | \Phi_{R'}^{\rm cl} \rangle
\langle   \Phi_{R'}^{\rm cl} | H_{QP} | \xi_E^c \rangle
\label{eq:16} 
\end{eqnarray}
where the
\begin{eqnarray}
\omega_R = G_P H_{PQ} \Phi_R^{\rm cl}
\label{eq:17} 
\end{eqnarray}
follow from the solutions of the Schr\"odinger equation 
\begin{eqnarray}
\sum_{c'} (H^{cc'} - E) \; \langle \xi_E^{c'\, *} |\omega_R \rangle = 
 \langle \xi_E^{c\, *} | H_{PQ} |  \Phi_R^{\rm cl} \rangle
\label{eq:18} 
\end{eqnarray}
with source term which connects the two sets $\{\Phi_R^{\rm cl}\}$ and
$\{\xi_E^c\}$ of wave functions.
With these coupling matrix elements  
\begin{eqnarray}
W_R^c =   \langle  \xi_E^{c\, *}  | H_{PQ} |\Phi_{R}^{\rm cl}  \rangle
=  \langle   \Phi_{R}^{\rm cl} | H_{QP} | \xi_E^{c}   \rangle
\label{eq:w1} 
\end{eqnarray}
between {\it discrete} states and 
scattering wave functions, it follows from (\ref{eq:heff}) 
\begin{eqnarray}
\langle \Phi_R^{\rm cl} | H^{\rm eff} |  \Phi_{R'}^{\rm cl} \rangle
 = \langle \Phi_R^{\rm cl} | H^{\rm cl} |  \Phi_{R'}^{\rm cl} \rangle 
+ \sum_{c=1}^K {\cal P} 
\int\limits_{\epsilon_{c}}^{\infty} dE' \;  
\frac{ W_R^c W_{R'}^c}{E-E'} - i \pi \; \sum_{c=1}^K  W_R^c W_{R'}^c 
\; .
\label{eq:heff1} 
\end{eqnarray}
The principal value integral does not vanish, in general.

With the eigenfunctions $\tilde \Phi_R$ and eigenvalues 
$\tilde {\cal E}_R = \tilde E_R - \frac{i}{2}\tilde \Gamma_R$ of  
$H^{\rm eff}$, the solution $\Psi_E^c$ of the Schr\"odinger equation in the
whole function space of discrete and scattering states reads
\begin{eqnarray}
\Psi_E^c 
=   \xi_E^c + \sum_{R} \tilde \Omega_R \;
\frac{\tilde W_R^c } 
{E- \tilde E_R + \frac{i}{2}\; \tilde \Gamma_R} 
\; .  
\label{eq:19} 
\end{eqnarray}
In order to identify the eigenvalues and eigenfunctions 
of $H^{\rm eff}$ with values of physical
relevance, the two subspaces have to be defined in an adequate manner. 
When the 
$P$ subspace contains all scattering states defined by their 
asymptotic behaviour and the $Q$ subspace 
is constructed from {\it all} the wave functions of the closed system
in a certain energy region (for details see \cite{baroho,ro91}), the values 
\begin{eqnarray}
\tilde W_R^c =  
 \langle \xi_E^{c \, *} | H_{PQ} | \tilde \Phi_{R} \rangle 
= \langle \tilde \Phi_{R}^*| H_{QP} | \xi_E^{c} \rangle 
\label{eq:w2} 
\end{eqnarray}
are the coupling coefficients between the {\it resonance} states 
and  scattering wave functions, while
the eigenvalues determine the energies
$\tilde E_R$ and widths $\tilde\Gamma_R$ 
of the resonance states. The 
\begin{eqnarray}
 \tilde \Omega_R = \tilde \Phi_R + \tilde \omega_R 
= (1 +  G_P H_{PQ}) \tilde \Phi_R
\label{eq:w3} 
\end{eqnarray}
are the wave functions
of the resonance states (with $\tilde \omega_R$ defined by Eq. (\ref {eq:17})
when $ \Phi_{R'}^{\rm cl}$ is replaced by $ \tilde \Phi_{R}$).

The Hamiltonian  $H^{\rm eff}$ is non-Hermitian since it is defined in a
subspace of the whole function space.
The left and  right eigenfunctions, $\tilde\Phi^{\rm lt}_R$ and 
$\tilde\Phi^{\rm rt}_R$, 
of a non-Hermitian matrix are different from one another. 
For a  symmetrical matrix, it follows 
\begin{equation}
\langle \tilde\Phi^*_R|\; H^{\rm eff} = \langle \tilde\Phi^*_R |\; 
{\cal E}_R  \qquad 
{\rm and} \qquad
H^{\rm eff} \; | \tilde\Phi_R \rangle = 
 {\cal E}_R\; | \tilde\Phi_R \rangle \; ,
\label{eq:orth5}
\end{equation}
see e.g. \cite{marost1,marost2,mudiisro,pegoro}.
Therefore,  $\tilde\Phi^{\rm lt}_R=\tilde\Phi^{\rm rt \; *}_R 
\equiv \tilde\Phi^ *_R $. 
The eigenfunctions of  $ H^{\rm eff}$ 
can be orthonormalized according to 
\begin{equation}
\langle\tilde\Phi^{\rm lt}_R|\tilde\Phi^{\rm rt}_{R'}\rangle = 
 \langle \tilde\Phi^*_R|\tilde\Phi_{R'}
\rangle =\delta_{RR'}
\label{eq:orth}
\end{equation}
where $\tilde\Phi^{\rm rt}_{R'} \equiv \tilde\Phi_{R'}$. 
Eq. (\ref{eq:orth}) provides
the bi-orthogonality relations
\begin{eqnarray}
\langle \tilde \Phi_R|\tilde\Phi_R \rangle =
\Re( \langle \tilde\Phi_R|\tilde\Phi_R \rangle) 
= \langle \tilde\Phi_{R'}|\tilde\Phi_{R'} \rangle \;
 \;  & ; & \; \; A_R \equiv   \langle \tilde \Phi_R|\tilde\Phi_R \rangle
\ge 1
\nonumber \\
 \langle \tilde\Phi_R|\tilde\Phi_{R'\ne R} \rangle  =  
i \; \Im (\langle \tilde\Phi_R|\tilde\Phi_{R'\ne R} \rangle )  
 =  -\langle \tilde\Phi_{R'\ne R}|\tilde\Phi_R \rangle 
\; \; & ; & \; \; B_R^{R'\ne R} \equiv  
| \langle\tilde \Phi_R|\tilde\Phi_{R'\ne R} \rangle| \ge 0 \; . 
  \label{eq:biorth}
\end{eqnarray}
Using the orthonormality condition (\ref{eq:orth}), it follows
\begin{eqnarray}
\tilde \Gamma_R = - 2   \, \Im \{ \langle \tilde \Phi_R^* | H^{\rm eff} |
\tilde \Phi_R \rangle \} = 2 \pi \sum_{c=1}^K (\tilde W_R^c )^2 
\label{eq:w3a} 
\end{eqnarray}
for the relation between the $\tilde \Gamma_R$ and $\tilde W_R^c$.
This relation holds also for overlapping resonance states. 
Here (\ref{eq:w3a}) holds, but $\tilde \Gamma_R  = 
(2 \pi/A_R) \sum |\tilde W_R^c |^2
\le 2 \pi  \sum |\tilde W_R^c |^2 $ according to (\ref{eq:biorth}).

It should be underlined here that the expression (\ref{eq:19}) is obtained by
rewriting the Schr\"odinger equation (\ref{eq:1}) with the only approximation
$P+Q=1$. The $\Psi_E^c$ and  $\xi_E^c$ as well as the $\tilde \Omega_R , \,  
\tilde E_R, \, \tilde \Gamma_R$ depend on the
energy $E$ of the system. The energies $E_R=\tilde E_R(E=E_R)$,  
widths $\Gamma_R = \tilde \Gamma_R(E=E_R)$ and wave functions
$\Omega_R = \tilde \Omega_R(E=E_R)$  
of the resonance states can be found by solving the corresponding 
fixed-point equations. Also the coupling matrix elements 
$\tilde W_R^c $
are complex and energy dependent functions, generally.
For numerical examples see \cite{drokplro}.

The expression (\ref{eq:19}) is solution of Eq. (\ref{eq:1}) 
independently of whether or not
the Hamilton operator $H$ contains two-body residual forces $V$.
It is $H=H_0+V$ e.g. in nuclear physics but  $H=H_0$
for quantum billiards. When $H=H_0+V$, it follows 
$ W_R^c = \langle  \Phi_{R}^{\rm cl} | V | \xi_E^c \rangle $ 
\cite{baroho,ro91}.
In the case of quantum billiards, the $W_R^c$ can be calculated by using
Neumann boundary conditions at the place of attachment of the lead 
to the cavity whereas Dirichlet boundary conditions are used at the 
boundary of the cavity (see  \cite{stm}). 
In this case, the $W_R^c$ are real, i.e. the principal value integral in 
(\ref{eq:heff1}) vanishes \cite{dittes}. The $\tilde W_R^c$ may be complex,
nevertheless \cite{stm}.

\subsection{The $S$ matrix}

The $S$ matrix is defined by the relation between the  incoming and 
outgoing waves in the asymptotic region. 
Its general form is
\begin{eqnarray} 
S_{cc'} = exp(2i\delta_c) \delta_{cc'} - 2 i \pi 
\langle \chi_E^{c'*} | V | \Psi_E^{c} \rangle 
\label{eq:smatr1}
\end{eqnarray}
where the $\chi_E^c$ are uncoupled scattering wave functions
obtained from  
\begin{eqnarray}  
\sum_{c'} (H_{0}^{cc'} - E) \; \chi_E^{c'} = 0 \; .
\label{eq:s1}
\end{eqnarray}
The $\Psi_E^{c}$ are given by Eq. (\ref{eq:19}).
When the two subspaces are defined consistently, 
Eq. (\ref{eq:smatr1}) can be written as
\begin{eqnarray} 
S_{cc'} = S_{cc'}^{(1)} -  S_{cc'}^{(2)} 
\label{eq:smatr2}
\end{eqnarray}
where 
\begin{eqnarray} 
 S_{cc'}^{(1)} = exp(2i\delta_c) \delta_{cc'} - 2 i \pi 
\langle \chi_E^{c'*} | V | \xi_E^{c} \rangle 
\label{eq:smatrdir}
\end{eqnarray}
is the smooth direct reaction part and 
\begin{eqnarray} 
 S_{cc'}^{(2)} =   2 i \pi  \sum_{R=1}^N
\langle \chi_E^{c'*} | V | 
  {\tilde \Omega}_R \rangle    \cdot
  \frac{\tilde W_{R}^{c}}
  {E - {\tilde E}_R + \frac{i}{2} {\tilde \Gamma}_R}  
\label{eq:smatres0}
\end{eqnarray}
is the resonance reaction part. In $S_{cc'}^{(1)}$,   $~V$
may be zero (no coupling between the channels).

Since Eq. (\ref{eq:w3}) between the resonance states $\tilde\Omega_R$
and the eigenfunctions $\tilde \Phi_R$ of $H^{\rm eff}$ is 
completely analogous to the Lippman-Schwinger equation 
\begin{eqnarray}
 \xi_E^c  = (1+G_P V)\; \chi_E^{c} 
\label{eq:lipp}
\end{eqnarray}
(which describes the relation between the two scattering wave functions
$\xi_E^c$  and $\chi_E^c$ with and without channel-channel coupling,
respectively),  one arrives  at \cite{ro91} 
\begin{eqnarray} 
\langle \chi_E^{c*} | V |\tilde \Omega_R \rangle
= \langle \xi_E^{c*} | V |\tilde \Phi_R \rangle 
= \tilde W_R^c 
\; .
\label{eq:coupl3}
\end{eqnarray}
When $V=0$, 
channel-channel coupling may appear due to the coupling of the scattering
wave functions via the $Q$ subspace (an effective Hamilton operator in the $P$
subspace can be derived analogously to the effective Hamilton operator 
$H^{\rm eff}$ in the $Q$ subspace, Eq. (\ref{eq:heff}), see \cite{ro91}).

Using this relation, the resonance part (\ref{eq:smatres0})
of the $S$ matrix reads
\begin{eqnarray} 
 S_{cc'}^{(2)} = 2 i \pi \; \sum_{R=1}^N
  \frac{\tilde W_{R}^{c} \; \tilde W_{R}^{c'}}
  {E - {\tilde E}_R + \frac{i}{2} {\tilde \Gamma}_R} \; .  
\label{eq:smatres}
\end{eqnarray}
Thus, the poles of the $S$ matrix are the eigenvalues 
$\tilde {\cal E}_R = \tilde E_R - \frac{i}{2}\; \tilde\Gamma_R$ 
of the Hamiltonian $ H^{\rm eff}$ at the energy $E=E_R$ (solutions of the
fixed-point equations). The numerator of Eq. (\ref{eq:smatres}) contains 
the squares $(\tilde \Phi_R )^2 $ of the eigenfunctions of $ H^{\rm eff}$ 
(and not the $|\tilde \Phi_R |^2 $) 
which are always finite in accordance with the 
normalization condition  (\ref{eq:orth}). Numerical results for 
strongly overlapping  resonances can be found in \cite{mudiisro}.

For overlapping
resonance states (i.e.  $W_R^c(E) W_{R'}^c(E) \ne 0$ 
at the energy $E$ for $R \ne R'$),
the $\tilde W$ may be very different from the $W$. 
Even when the $W$ are real, the $\tilde W$ may be complex
since the eigenfunctions  of $H^{\rm eff}$ are complex.
The coupling strength of the system to the
continuum of channel wave functions
is given by the sum of the imaginary parts of the diagonal matrix elements or
of the eigenvalues of $H^{\rm eff}$,
\begin{eqnarray}
 \sum_R \Gamma_R =  \sum_R \tilde \Gamma_R = 
2\pi \sum_{Rc} \tilde W_R^{c} \tilde W_{R}^c = 
2\pi \sum_{Rc} W_R^c W_{R}^c 
\label{eq:wsum} 
\end{eqnarray}
where further Eq. (\ref{eq:w3a}) is used.
All redistributions taking place in the system under the influence of 
a certain parameter must obey the sum rule (\ref{eq:wsum}). According to this
rule, resonance trapping may appear when the resonance states overlap,
\begin{eqnarray}
\sum_{R=1}^N \tilde \Gamma_R \approx \sum_{R=1}^M \tilde \Gamma_R \;  
\; \; ; \quad
\sum_{R=M+1}^N \tilde \Gamma_R \approx 0 \; .
\label{eq:trap} 
\end{eqnarray}
It means that, under certain conditions, $N-M$ resonance states 
may  decouple from the continuum of scattering states,
i.e. they may be  {\it trapped}
by $M$ states. For numerical results on nuclei see \cite{ro91}
and on open quantum billiards see \cite{napirose}.
Studying the system by means of its coupling to the environment 
(described by the coupling matrix elements $\tilde W_R^c$) will give, in such
a case, information either on the $N-M$ long-lived states on the background  
of the $M$ short-lived states or on the $M$ short-lived states with
fluctuations arising from the $N-M$ long-lived states.

This behaviour induced by the imaginary part of the non-diagonal matrix
elements of (\ref{eq:heff1}) differs from that induced by the real part. While
the imaginary part causes the formation of structures with different time
scales (as discussed above), the real part causes equilibrium in time 
(approaching of the decay widths). The first case is accompanied by an
approaching of the states in energy ({\it clustering of levels}) while the
second case is accompanied by  {\it level repulsion in energy}. 
Numerical examples
can be found in \cite{marost1,marost2} for atoms and in \cite{ropepise} 
for quantum
billiards. The interplay between the real and imaginary parts of the
non-diagonal matrix elements of (\ref{eq:heff1}) characterizes the dynamics of
open quantum systems.

Eq. (\ref{eq:smatres}) coincides formally with the standard
form of the resonance part of the $S$ matrix.
It should be underlined, however, that the 
$\tilde W_{R}^{c}, ~\tilde E_R$, and  $\tilde \Gamma_R$ 
are not parameters (as in the standard form, see \cite{mawei}),
but {\it energy dependent functions} which can be calculated. 
Eq.  (\ref{eq:smatres}) holds also for overlapping resonances.
Due to the energy dependence of the $\tilde \Gamma_R$, 
the line shape of the resonances differs from a 
Breit-Wigner shape, as a rule, even without any interferences with the 
direct reaction part. Some numerical examples are discussed in 
\cite{marost1,ro91,mudiisro}.

\section{Branch points in the complex plane}

Eq. (\ref{eq:smatres}) gives the $S$ matrix for isolated as well as for
overlapping resonance states. The extreme case of overlapping 
corresponds to the double pole of the $S$ matrix at which the eigenvalues
$\tilde{\cal E}_{1,2}$
of two resonance states are equal. Let us illustrate this case 
by means of the
complex two-by-two Hamiltonian matrix 
 \begin{eqnarray}
{\cal H} =
 \left(
\begin{array}{cc}
 e_1(a) & 0 \\
0  &   e_2(a)
\end{array}
\right) -
 \left(
\begin{array}{cc}
 \frac{i}{2}\gamma_1(a) &  \omega\\
 \omega &  \frac{i}{2} \gamma_2(a)
\end{array}
\right)  \; .
\label{eq:matr1}
\end{eqnarray}
The unperturbed energies $e_k$ and widths $\gamma_k$
($k=1,2$) of the two states  depend  on 
the parameter $a$ to be tuned in such a manner that  
the two states may cross in energy (and/or width) when $\omega = 0$.
The two states interact only via the non-diagonal matrix elements 
$\omega $ which  may be complex, in general, see Eq. (\ref{eq:heff1}).
In the following, we consider  real $\omega$ and
$\gamma_k$ independent of $a$.

The eigenvalues of  ${\cal H}$     are
\begin{equation}
{\cal E}_{i,j} \equiv E_{i,j} - \frac{i}{2}  \Gamma_{i,j} =
\frac{\epsilon_1 + \epsilon_2}{2} \pm \frac{1}{2} \;
\sqrt{(\epsilon_1 - \epsilon_2)^2 + 4 \omega^2}
\label{eq:avoi}
\end{equation}
with $i, j = 1, 2$ and
 $\epsilon_k \equiv e_k - \frac{i}{2} \; \gamma_k \; \; (k=1,2)$.
According to Eq. (\ref{eq:avoi}),  two interacting discrete states 
(with $\gamma_k = 0$) avoid
always crossing since $\omega$ and $(\epsilon_1 - \epsilon_2)$
are real in this case. 
Eq. (\ref{eq:avoi}) shows  also that  resonance states
with non-vanishing widths $\gamma_k$ avoid 
mostly crossing since
\begin{eqnarray}
  F(a,\omega) \equiv  (\epsilon_1 - \epsilon_2)^2 + 4 \omega^2 
\label{eq:sqrtf}
\end{eqnarray}
is different from zero for all $a$, as a rule. Only when  
$ F(a,\omega) = 0$ at $a=a^{\rm cr}$ (and $\omega=\omega^{\rm cr}$),
the states cross, i.e. ${\cal E}_1 = {\cal E}_2$. 
In such a case, the $S$ matrix has a double pole, see e.g. \cite{newton}.

It can further be seen from Eq. (\ref{eq:avoi}) that
the crossing points are branch points in the complex plane. 
The  branch point is determined by 
the  values $(\omega )^2$ and $(\epsilon_1 - \epsilon_2)^2$
but not by the signs of these values.
According to Eq. (\ref{eq:avoi}),   it lies 
in the complex plane at the point
$ X \equiv (1/2) \{\epsilon_1(a^{\rm cr}) + \epsilon_2(a^{\rm cr})\} $.
According to Eq.  (\ref{eq:smatres}), 
it is a double pole of the $S$ matrix.

The eigenfunctions $\Phi_i $  can be represented in the set of
basic wave functions $\Phi_i^0$ of the  unperturbed matrix 
corresponding to  $\omega =0$,
\begin{equation}
\Phi_i=\sum b_{ij} \Phi_j^0 \; .
\label{eq:mix}
\end{equation}
 In the critical region of avoided crossing, the
eigenfunctions are mixed: $|b_{ii}|=|b_{jj}| \ne  1$  and 
$b_{ij}=-b_{ji} \ne 0$ for $i\ne j$. The $b_{ij}$ are normalized according to 
Eq. (\ref{eq:orth}).

The Schr\"odinger equation with the Hamiltonian ${\cal H}$ 
can be rewritten as
\begin{eqnarray}
({\cal H}_0 - {\cal E}_i) |\Phi_i\rangle & = & 
 \left(
\begin{array}{cc}
0 &  \omega \\
 \omega  &  0
\end{array}
\right) |\Phi_i\rangle
 \equiv
 \; W \;
 |\Phi_i\rangle 
  \nonumber \\
& = &  \sum_{k=1,2} \langle \Phi_k| W |\Phi_i\rangle \sum_{m=1,2}
\langle \Phi_k| \Phi_m\rangle |\Phi_m\rangle \; .
\label{eq:nls1} 
\end{eqnarray}
Eq. (\ref{eq:nls1}) is a   Schr\"odinger equation 
 with the Hamiltonian 
${\cal H}_0$ of the unperturbed system (corresponding to $\omega=0$) 
and a source term which is related directly to the
bi-orthogonality of the eigenfunctions   of the Hamiltonian ${\cal H}$. 
It vanishes with $|\Phi_i|^2 \equiv \langle\Phi_i |\Phi_i\rangle \to 1$ 
and $ |\langle\Phi_i |\Phi_{j\ne i} \rangle | \to 0$. 
Using Eq. (\ref{eq:biorth}), the rhs of Eq. (\ref{eq:nls1}) reads
\begin{eqnarray}
W |\Phi_i \rangle = W^{1i} \Big(A |\Phi_1\rangle + i B |\Phi_2\rangle \Big) + 
W^{2i} \Big(A |\Phi_2 \rangle - i B |\Phi_1 \rangle \Big)
\label{eq:wab}
\end{eqnarray}
with $ W^{ki} \equiv 
\langle \Phi_k | W | \Phi_i \rangle ; \; k=1,2 $.
At the branch point, the complex energies are equal, ${\cal E}_1^{bp} =
{\cal E}_2^{bp} $, and 
\begin{eqnarray}
 W^{11}_{bp} \Big(A |\Phi_1^{bp}\rangle & + & i B |\Phi_2^{bp}\rangle \Big) + 
W^{21}_{bp} \Big(A |\Phi_2^{bp} \rangle - i B |\Phi_1^{bp} \rangle \Big)  
\nonumber \\ 
& = & W^{12}_{bp} \Big(A |\Phi_1^{bp}\rangle + i B |\Phi_2^{bp}\rangle \Big) + 
W^{22}_{bp} \Big(A |\Phi_2^{bp} \rangle - i B |\Phi_1^{bp} \rangle \Big)
\nonumber \\
& = & W^{11}_{bp} \Big(A |\Phi_2^{bp}\rangle - i B |\Phi_1^{bp}\rangle \Big) + 
W^{21}_{bp} \Big(A |\Phi_1^{bp} \rangle + i B |\Phi_2^{bp} \rangle \Big)
\label{eq:wab2}
\end{eqnarray}
where the relations $W^{11} = W^{22}\; ; \; \, W^{12} = W^{21}$ are
used. This gives 
\begin{eqnarray}
(A+ i B) |\Phi_1^{bp}\rangle = (A - i B) |\Phi_2^{bp}\rangle   
\label{eq:wab3}
\end{eqnarray}
and finally
\begin{eqnarray}
|\Phi_i^{bp}\rangle & = & \Big(1-2\frac{B^2}{A^2 + B^2} \pm 2 i \frac{AB}{A^2 +
B^2} \Big)|\Phi_{j\ne i}^{bp}\rangle 
\nonumber \\
& \to &
\pm \; i\; | \Phi_{j\ne i}^{bp}\rangle
\label{eq:wfexch}
\end{eqnarray}
by using 
the bi-orthogonality relations (\ref{eq:biorth}) 
\begin{eqnarray}
|\langle \Phi_i^{bp}|\Phi_i^{bp} \rangle| \to \infty \qquad 
|\langle \Phi_i^{bp}|\Phi_{j\ne i}^{bp} \rangle| \to \infty 
\label{eq:b1}
\end{eqnarray}
at the branch point. The relation (\ref{eq:wfexch})  between the
two wave functions at the branch point in the complex plane
has been proven in numerical calculations for the hydrogen atom with 
a realistic Hamiltonian \cite{marost2}.
Note, that the condition  (\ref{eq:orth}) 
\begin{eqnarray}
\langle \Phi_i^{bp\, *}|\Phi_j^{bp} \rangle =\delta_{ij} \; .
\label{eq:b2}
\end{eqnarray}
is fulfilled also at the branch point in the complex plane. This
is achieved since
the difference between two infinitely large values may be 0
(for $i\ne j$) or 1 (for $i=j$). 
Numerical examples for 
the values $ \langle \Phi_i|\Phi_j \rangle $ with $i=j$ as well as 
with $i\ne j$ 
can be found in \cite{mudiisro,pegoro,slaving,pepirose,seromupepi}.

Thus, the condition (\ref{eq:orth}) is fulfilled
in the whole function space without any exception. 
According to Eq. (\ref{eq:wfexch}), the two wave functions can be 
exchanged by means of interferences
between $A$ and $B$, i.e. by means of the source term in the Schr\"odinger
Eq. (\ref{eq:nls1}). When the wave functions are
exchanged, $\Re (b_{11}) = \Re (b _{22})$ as well as $\Re (b_{12}) 
= - \Re (b_{21})$ do
not change their signs while $\Im (b_{11}) = \Im (b_{22})$ and $\Im (b_{12}) 
= - \Im (b_{21})$ jump between $-\infty$ and $+\infty$ at the branch point.

Note that Eq. (\ref{eq:wfexch}) differs from the relation $\Phi_1^{bp} =
\pm \Phi_2^{bp}$ used  in the literature for the definition of the
exceptional point \cite{hemuro}.
With $\Phi_1^{bp} = \pm \,\Phi_2^{bp}$, neither 
 $\langle \Phi_i^{bp}|\Phi_j^{bp} \rangle = \delta_{ij}$ nor
$\langle \Phi_i^{bp\;*}|\Phi_j^{bp} \rangle = \delta_{ij}$ can  be fulfilled. 
With $\Phi_1^{bp} = \pm \, i \, \Phi_2^{bp}$ however,
the orthogonality relations (\ref{eq:orth}) are fulfilled in
the whole function space, without any exception.

The  effects arising from the source term 
in Eq. (\ref{eq:nls1}) play the decisive role 
in the dynamics of many-level
quantum systems caused by avoided level crossings.
This will  be illustrated 
in the following section by means of numerical results.
The effects appear everywhere in the complex plane when only 
$ |\Phi_i|^2 \ne 1, \;|\langle \Phi_i |
\Phi_{j\ne i} \rangle| \ne 0 $. They appear also in the function space 
of discrete states where the $\Phi_i$ are real, due to the analyticity
of the wave functions and their continuation 
into the function space of discrete states.

\section{Numerical results}

\begin{figure}
\begin{minipage}[tl]{7.5cm}
\psfig{file=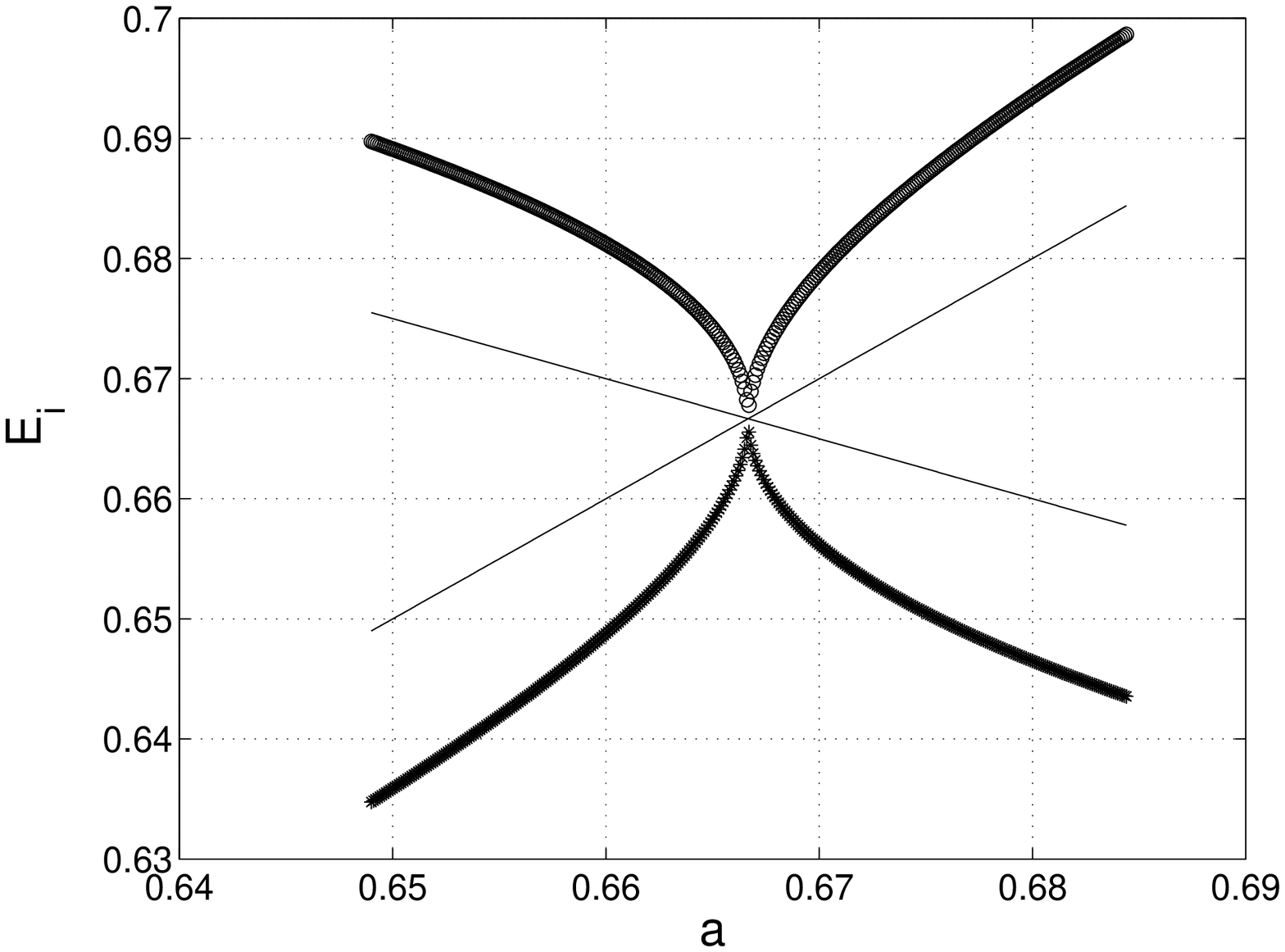,width=7.5cm}
\end{minipage}
\begin{minipage}[tr]{7.5cm}
\psfig{file=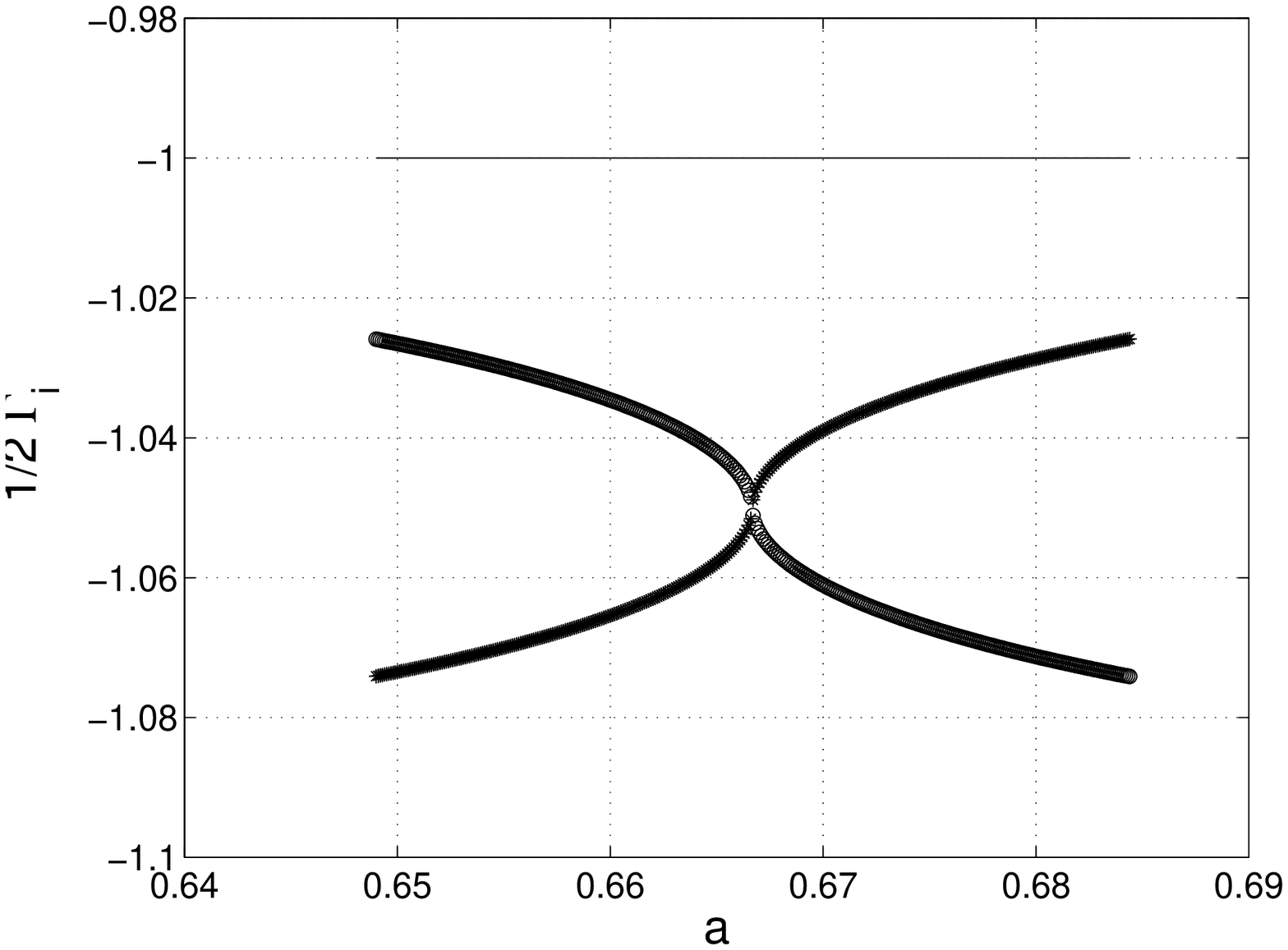,width=7.5cm}
\end{minipage}
\hspace*{.4cm}
\begin{minipage}[bl]{7.5cm}
\psfig{file=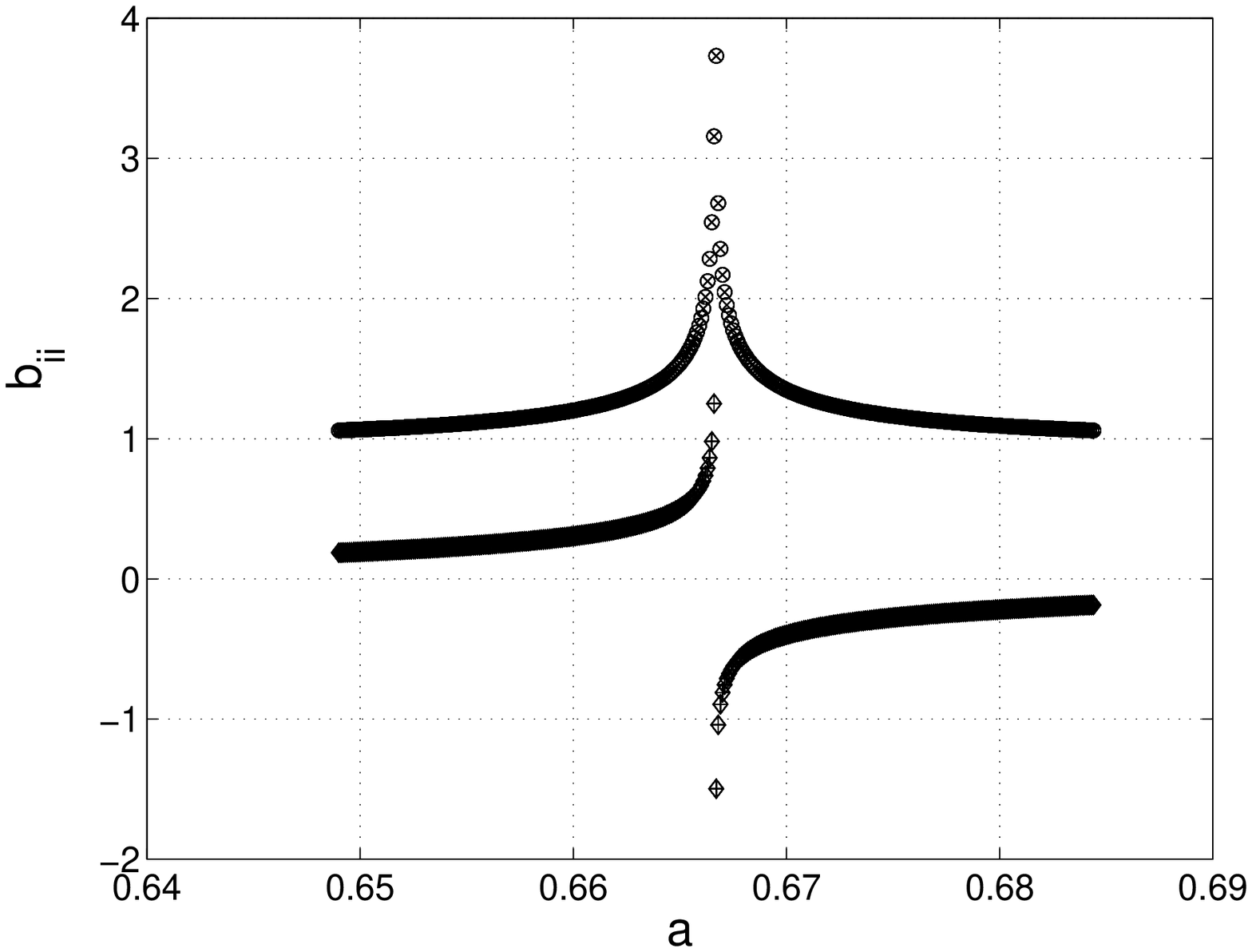,width=7.5cm}
\end{minipage}
\hspace*{.6cm}
\begin{minipage}[br]{7.5cm}
\psfig{file=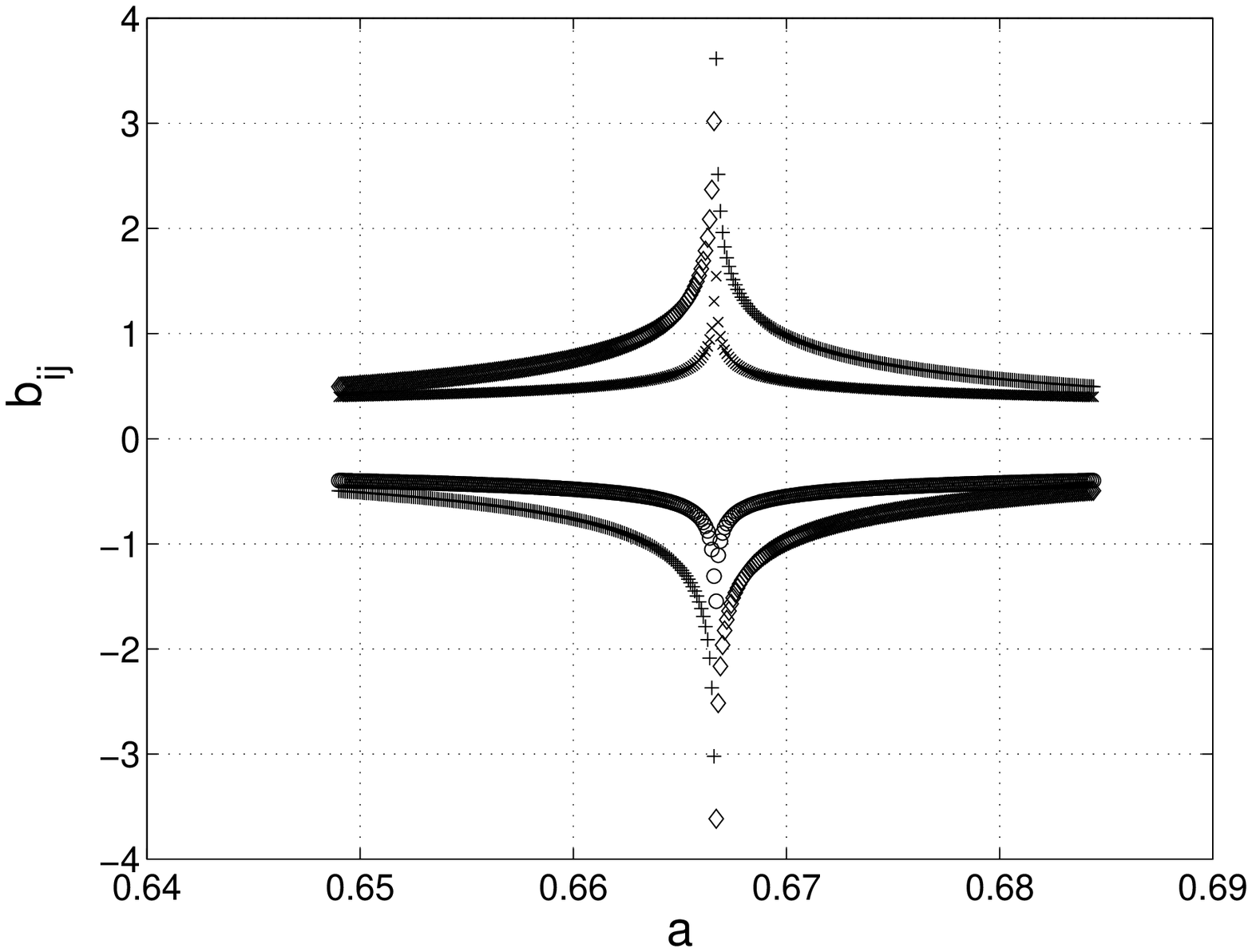,width=7.5cm}
\end{minipage}
\vspace*{.5cm}
\caption{The energies $E_i$ (top left) and widths
$\Gamma_i /2$  (top right) of   
the two eigenstates of the matrix (\ref{eq:matr1})
as a function of the parameter $a$. 
The thin lines give the energies $E_i$ and widths $\Gamma_i /2$ of the states 
at $\omega =0$.
The lower part of the figure shows the
 coefficients $b_{ii}$ (bottom left) 
 and  $b_{ij\ne i}$ (bottom right) defined by Eq. (\ref{eq:mix}). 
The  x and o denote 
the $\Re (b_{ij})$  while the  $\Im (b_{ij})$
are  denoted by + and $\diamond$. 
$e_1=1-a/2; \; e_2=a; \; \gamma_1 /2=1.0; \;
\gamma_2 /2 =  1.1$ and $\omega = 0.05$.
}
\label{fig:basic1}
\end{figure}

The numerical results obtained by diagonalizing the matrix (\ref{eq:matr1})
are shown in Figs. \ref{fig:basic1} to \ref{fig:avoi3}. The
$E_i$ and $\Gamma_i$ are in  units of $a$ chosen arbitrarily, 
and the $b_{ij}$ are dimensionless.
In all cases $e_1=1-a/2, \; e_2=a $ and $\omega = 0.05$. The $\gamma_i$ do
not depend on the tuning parameter $a$. The relation between them is 
$\gamma_2 = 1.1 \gamma_1$.  At $a = a^{\rm cr}=2/3$, 
the two levels cross when
unperturbed (i.e. $\omega = 0$) and avoid crossing, as a rule, when 
the  interaction $\omega$ is different from zero. 
Here $e_1 = e_2 \equiv e_{i=1,2}^{\rm cr} = 2/3$ and
\begin{eqnarray}
F^{\rm cr}(a,\omega) = \Re(F^{\rm cr}(a,\omega)) =   
4 \omega^2 - (1/4)\; (\gamma_i - \gamma_j)^2  
\label{eq:fcr}
\end{eqnarray}
according to Eqs. (\ref{eq:avoi}) and (\ref{eq:sqrtf}).
The $F^{\rm cr}(a,\omega)$ may be positive or negative. Thus, {\it either} the
widths $\Gamma_{i,j}$ {\it or} the energies $E_{i,j}$  
cross freely at $a^{\rm cr}$, but not both. The only exception
occurs when the $S$ matrix has a double pole at $a^{\rm cr}$, i.e.
when $\gamma_1 /2 = \gamma_1^{\rm cr} /2 =  
1.0, \; \gamma_2 /2 = \gamma_2^{\rm cr} /2 = 1.1$.  Here, 
$ F^{\rm cr}(a,\omega) = 0$
and  the two resonance states cross in spite of $\omega \ne 0$. 
According to Eq. (\ref{eq:avoi}),
the double pole of the $S$ matrix is a branch point in the complex plane.

In Fig. \ref{fig:basic1}, the energies $E_{i,j}$, 
widths $\Gamma_{i,j}$ and wave
functions $b_{ij}$ of the two states are shown as a function of the parameter 
$a$ in the very neighbourhood of the branch point.   
Approaching the branch point at $a^{\rm cr}$, 
$\; |\Re (b_{ij})| \to \infty$  and $|\Im (b_{ij})| \to
\infty$. While  $\Re (b_{ij})$ does not change its sign by crossing the 
critical
value $a^{\rm cr}$, the phase of  $\Im (b_{ij})$ jumps from $\pm$ to $\mp$. 
The  orthogonality relations (\ref{eq:orth}) are fulfilled for all $a$
including the critical value $a^{\rm cr}$.

Fig. \ref{fig:tra1} 
shows the energies $E_{i,j}$ and widths $\Gamma_{i,j}$ of the two states for
values of $\gamma_{i,j}$ just above and below the critical values 
$\gamma_{i,j}^{\rm
cr}$ as well as for $\gamma_{i,j} = 0$, i.e. for discrete states.  
According to Eq. (\ref{eq:fcr}),
either the energy trajectories   or the trajectories
of the widths avoid crossing at the critical value $a^{\rm cr}$
(since the condition for the appearance of a double pole
of the $S$ matrix is not fulfilled).  
It is exactly this  behaviour of the trajectories 
which can be seen in Fig. \ref{fig:tra1}:
\begin{enumerate}
\item[--]
When $\gamma_i >
\gamma_i^{\rm cr}$, the widths of the two states approach each other 
near $a^{\rm cr}$ but the width of one of the states remains always larger than
the width of the other one. 
The two states cross freely in energy, and the wave
functions are {\it not}
exchanged after crossing the critical value  $a^{\rm cr}$. 
\item[--]
The situation is
completely different when $\gamma_i <
\gamma_i^{\rm cr}$. In this case, the states avoid crossing in energy  
while their widths  cross freely. After crossing the critical value  $a^{\rm
cr}$, the wave functions of the two states are exchanged. An exchange  of the
wave functions takes place also in the case of discrete states 
($\gamma_i = 0$). This latter
result is well known as Landau-Zener effect. It is directly related to the 
branch point
in the complex plane at $a^{\rm cr}$ as can be seen from Fig.
\ref{fig:tra1}. 
\end{enumerate}

The wave functions $b_{ij}$ are shown in Fig.  \ref{fig:tra3}.
The states are mixed (i.e.  $|b_{ii}| \ne 1$ and $b_{ij\ne i} \ne 0$)
 in all cases in the neighbourhood of   $a^{\rm cr}$.  
In the case without exchange of the wave functions, $\Re (b_{ij})$ as well as 
$\Im (b_{ij})$ behave smoothly at $a^{\rm cr}$ while this is true only for 
$\Re (b_{ij})$ in the case with exchange of the wave functions. In this case, 
$\Im (b_{ij})$ jumps 
from a certain finite value $y$ to $-y$ at $a^{\rm cr}$.
Since the $\Im (b_{ij})$ of discrete states are zero, a jump in the 
 $\Im (b_{ij})$ can not appear in this case. 
The $\Re (b_{ij})$, however, show a dependence
on $a$ which is very similar to that of resonance states  with exchange
of the wave functions ($\gamma_i < \gamma_i^{\rm cr}$). 

In order to trace the influence of the branch point in the complex plane
onto the mixing of discrete states, the
differences  $\delta = |b_{ii}|^2 - |b_{ij\ne i}|^2$ and the values
 $|b_{ij}|^2$
are shown in Figs. \ref{fig:avoi1} and \ref{fig:avoi2} for different values
$\gamma_i$ from $\gamma_i > \gamma_i^{\rm cr}$ to $\gamma_i=0$.
Most interesting is the change 
of the value $\delta$ from 1 to 0 at  $\gamma_i^{\rm cr}$. The relation
$|b_{ii}|^2 = |b_{ij\ne i}|^2 $ at  $\gamma < \gamma_i^{\rm cr}$ is the 
result from interference processes. It holds also at  $\gamma_i =0$, 
i.e. for discrete states. In this case,  
$|b_{ii}|^2 = |b_{ij\ne i}|^2 = 0.5$ at $a^{\rm cr}$. 

The values $A$ and $B$ characterizing the bi-orthogonality of the two 
wave functions are shown in Fig. \ref{fig:avoi3} for the same values of 
$\gamma_i$ as in Figs. \ref{fig:avoi1} and \ref{fig:avoi2}.
The $A$ and $B$ are similar for $\gamma_i - \gamma_i^{\rm cr} = \pm \Delta$ 
as long as   $\Delta$ is small. They approach 
 $A \to 1$ and $ B \to 0$ for $\gamma_i \to 0$.

In Fig.  \ref{fig:four}, the energies $E_i$ and mixing coefficients 
 $|b_{ij}|^2$ are shown for illustration for four discrete states with three
 neighboured avoided crossings as a function of $a$. 
In analogy to (\ref{eq:matr1}), the matrix is 
 \begin{eqnarray}
{\cal H}^{(4)} =
 \left(
\begin{array}{cccc}
 e_1(a) & 0 & 0 & 0\\
0  &   e_2(a) & 0 & 0 \\
0  &  0 & e_3(a) & 0 \\ 
0  & 0 & 0 &  e_4(a) 
\end{array}
\right) -
 \left(
\begin{array}{cccc}
 0 &  \omega_{12} & \omega_{13} & \omega_{14} \\
  \omega_{21} & 0 & \omega_{23} & \omega_{24} \\
  \omega_{31} & \omega_{32} & 0 & \omega_{34} \\
 \omega_{41} & \omega_{42} &  \omega_{43} & 0\\
\end{array}
\right)  \; .
\label{eq:matr4}
\end{eqnarray}
The mixing in the eigenfunctions of ${\cal H}$ which is caused by the avoided
crossings remains, at high level density,
  at all values of the parameter $a$.  It is the result of
 complicated interference processes. This can be seen best by comparing
 the two pictures with four interacting states (top and middle 
in Fig.  \ref{fig:four}) with those of only two
 interacting states (bottom of Fig.  \ref{fig:four} and bottom right of
 Fig. \ref{fig:avoi2}). Fig. \ref{fig:four} bottom shows the large region
of the $a$ values around $a^{\rm cr}$ for which the two wave functions remain
mixed: $|b_{ii}|^2 \to 1$ and $|b_{ij\ne i}|^2 \to 0$ for $a \to a^l$ with 
$|a^l - a^{\rm cr}_{34}| \gg |a^{\rm cr}_{i4} - a^{\rm cr}_{34}| \; ; \; 
i=1,2$.
The avoided crossings between neighboured states do, therefore, not occur
between states with pure wave functions and it is  impossible  to
identify the  $|b_{ij}|^2$ unequivocally (Figs.  \ref{fig:four} right top and
middle). These avoided crossings  
are caused by branch points which {\it overlap} in the complex
plane while the avoided crossings  
considered in Figs. 
\ref{fig:basic1} to \ref{fig:avoi3} and Fig. \ref{fig:four} bottom
correspond to {\it isolated} branch points in the complex plane.

\section{Discussion of the results}

Most  calculations  represented in the present paper 
are performed for two states which cross or avoid crossing
under the influence of an interaction $\omega$ which is real. 
A general feature appearing in all the results is the  repulsion
of the levels in energy (except in the very neighbourhood of
$ a^{\rm cr}$ when $
\gamma_i \ge \gamma_i^{\rm cr}$, i.e. $
F^{\rm cr}(a,\omega) \le 0 $). 
This result follows analytically from the
eigenvalue equation  (\ref{eq:avoi}). It holds quite generally for real
$\omega$  as shown by means of
the spectra of microwave cavities   \cite{ropepise}
and laser-induced continuum structures in atoms \cite{marost1}.
The level repulsion in energy is accompanied by an approaching of  the
lifetimes (widths)  of the  states.

The sign of $F^{\rm cr}(a,\omega)$, Eq. (\ref{eq:fcr}),
 is decisive whether or not the states will be exchanged at
 the critical value   $a^{\rm cr}$ of the tuning parameter.
 When  $\omega$ is real and so small that $F^{\rm cr} < 0$ and the
 difference of the  widths $\Gamma_i - \Gamma_{j\ne i}$ 
is different from zero at 
 $a^{\rm cr}$ then the states  {\it will not be exchanged}
and the energy trajectories cross freely. If, however, 
$F^{\rm cr} > 0$ and
$\Gamma_i = \Gamma_{j\ne i}$ at $a^{\rm cr}$, the states {\it will 
be exchanged}
and the energy trajectories avoid crossing.

The exchange of the wave functions can be traced back to 
the branch point in the complex plane where the $S$ matrix 
has a double pole and $\Phi_i \to \pm \; i \; \Phi_j$
according to Eq. (\ref{eq:wfexch})  (and
calculations for a  realistic case  \cite{marost2}).
Here,  the real as well as the imaginary parts of the components
of the wave functions increase up to an infinite value and 
$\langle \Phi_i^*| \Phi_{j} \rangle $ is the difference between 
two infintely large values. Thus, the 
orthogonality relation $\langle \Phi_i^*| \Phi_{j\ne i} \rangle = 0 $
and the normalization condition   
 $\langle \Phi_i^*| \Phi_{i} \rangle = 1$ can not be distinguished.
This makes possible the exchange of the two wave functions.

\begin{figure}
\hspace{-1.8cm}
\begin{minipage}[tl]{7.5cm}
\psfig{file=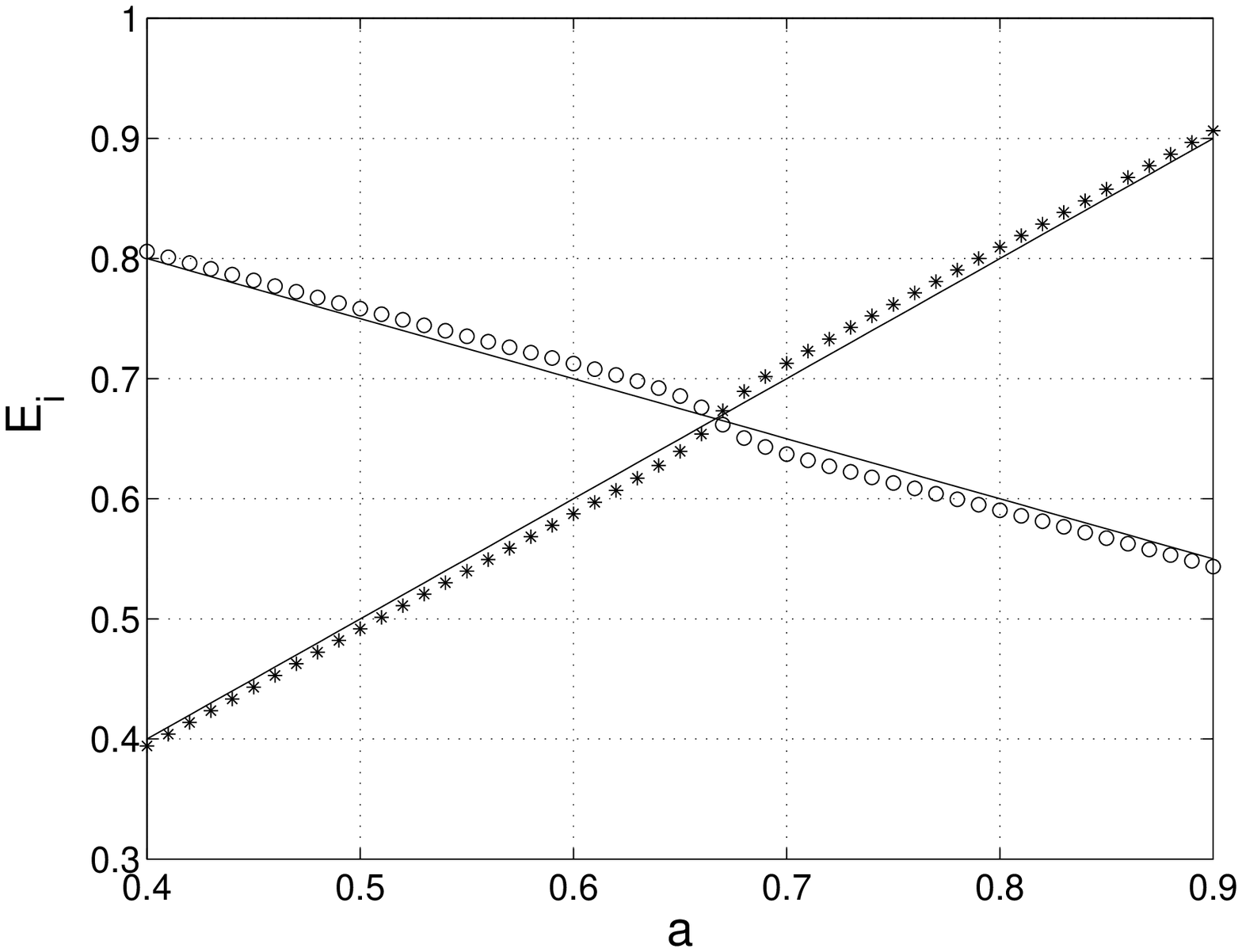,width=7.5cm}
\end{minipage}
\begin{minipage}[tr]{7.5cm}
\psfig{file=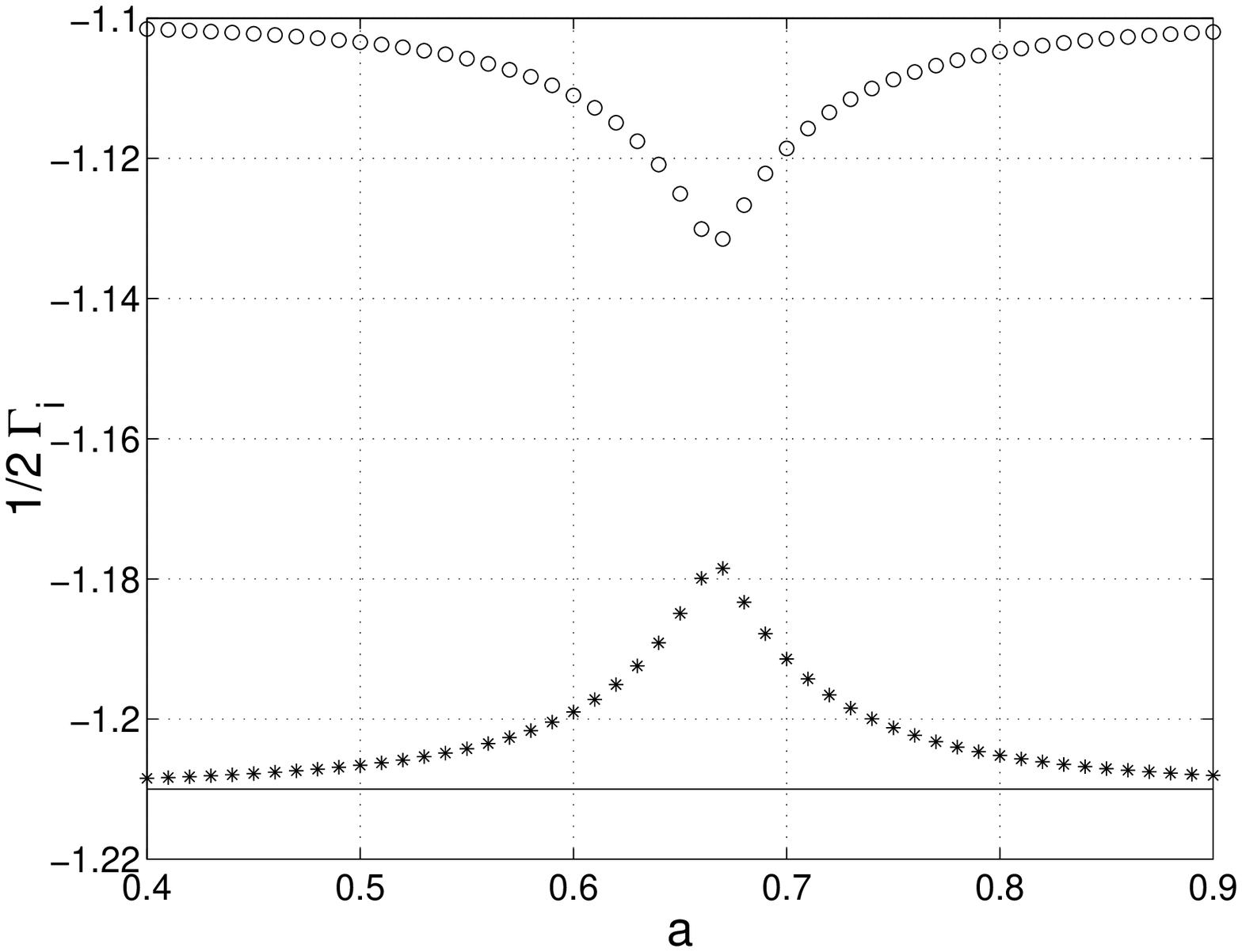,width=7.5cm}
\end{minipage}
\begin{minipage}[ml]{7.5cm}
\psfig{file=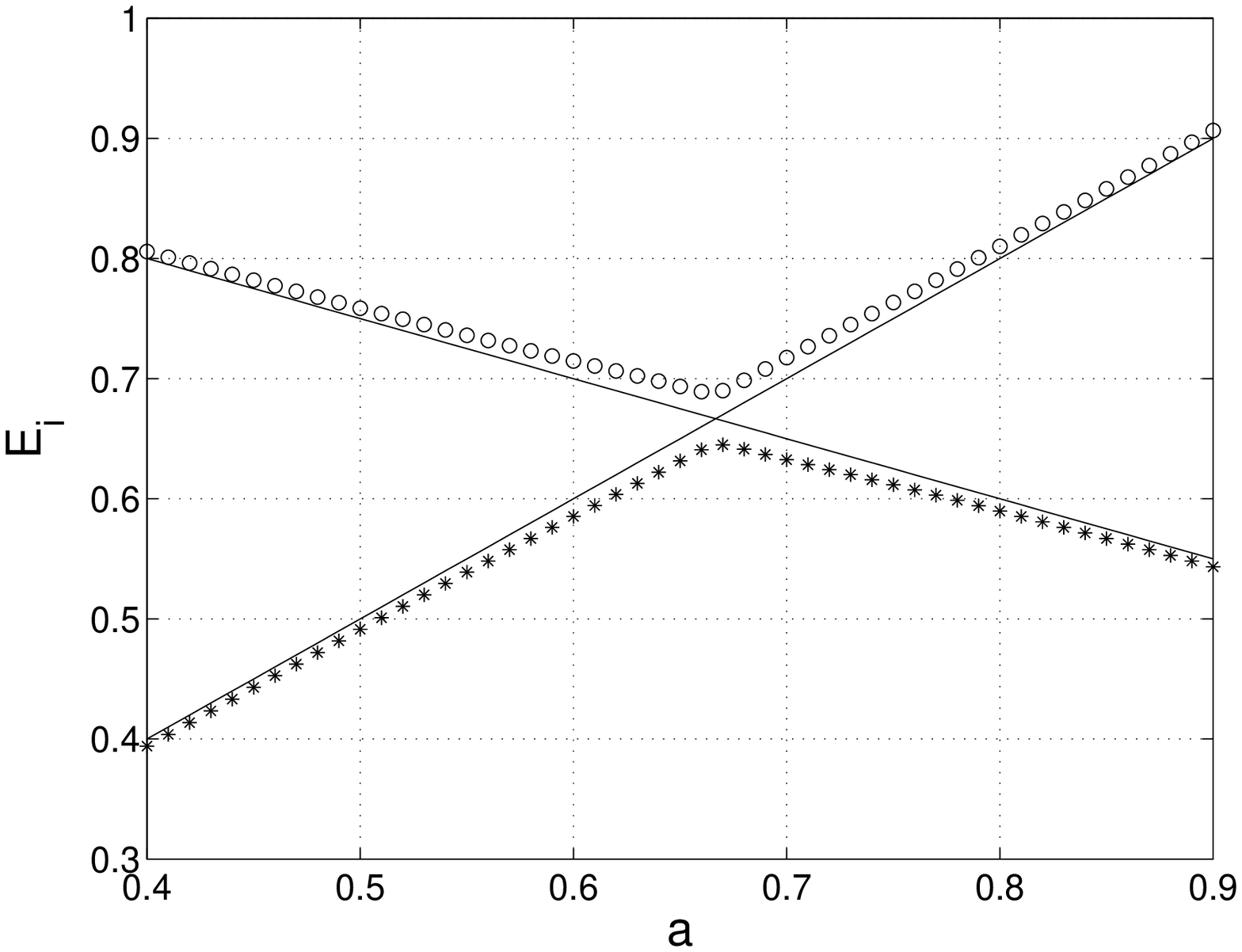,width=7.5cm}
\end{minipage}
\begin{minipage}[mr]{7.5cm}
\psfig{file=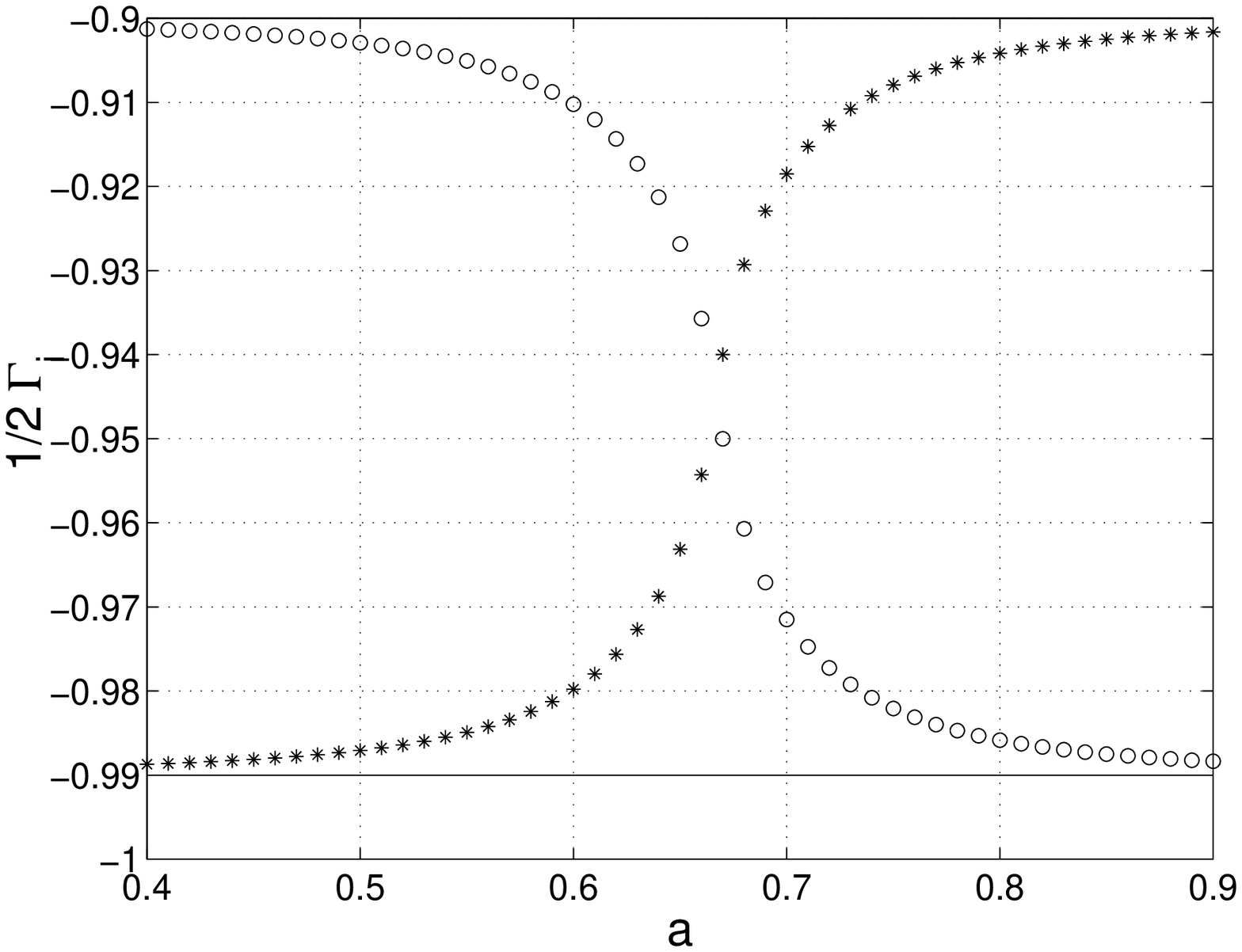,width=7.5cm}
\end{minipage}
\begin{minipage}[bl]{7.5cm}
\psfig{file=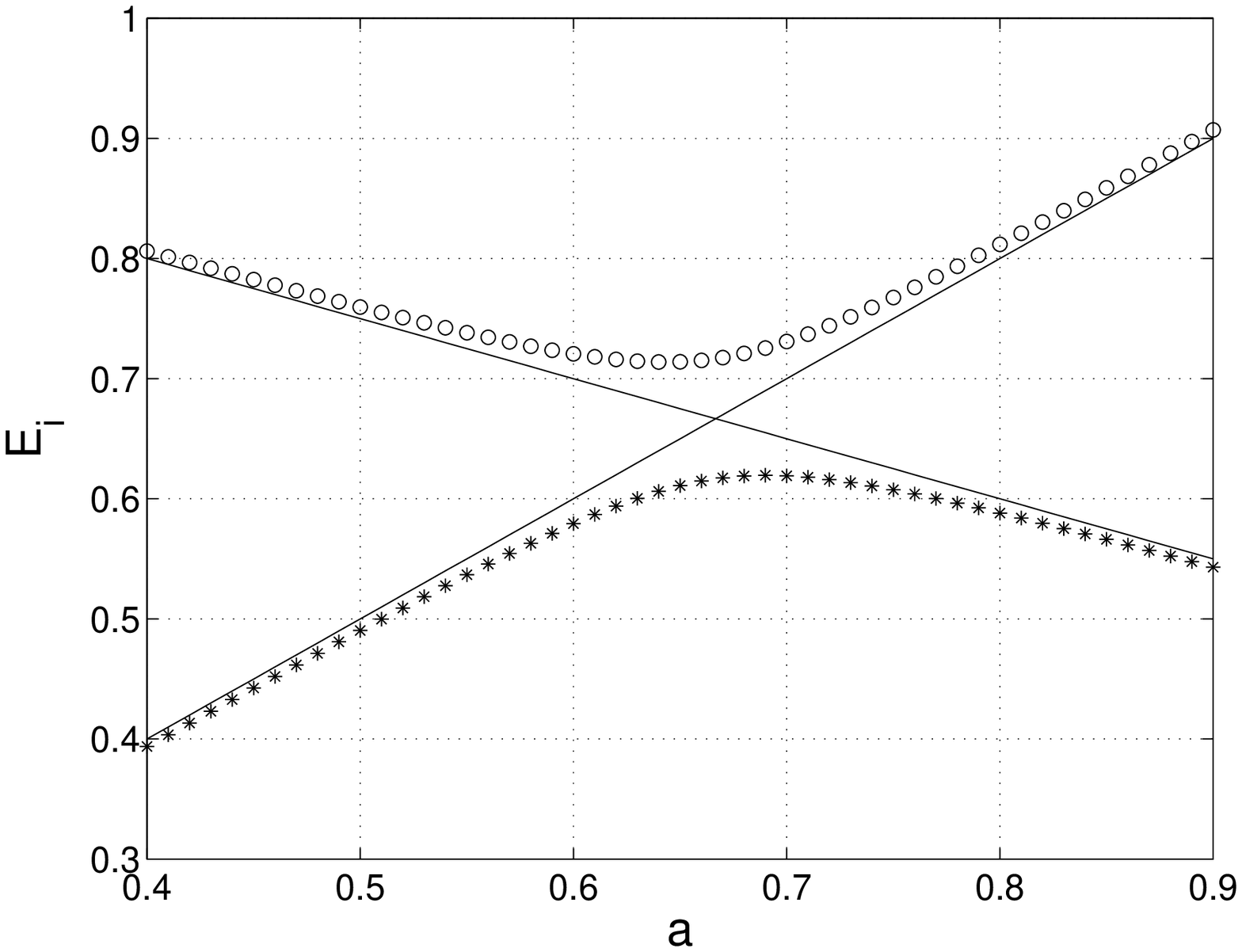,width=7.5cm}
\end{minipage}
\hspace{1.0cm}
\begin{minipage}[br]{7.5cm}
\psfig{file=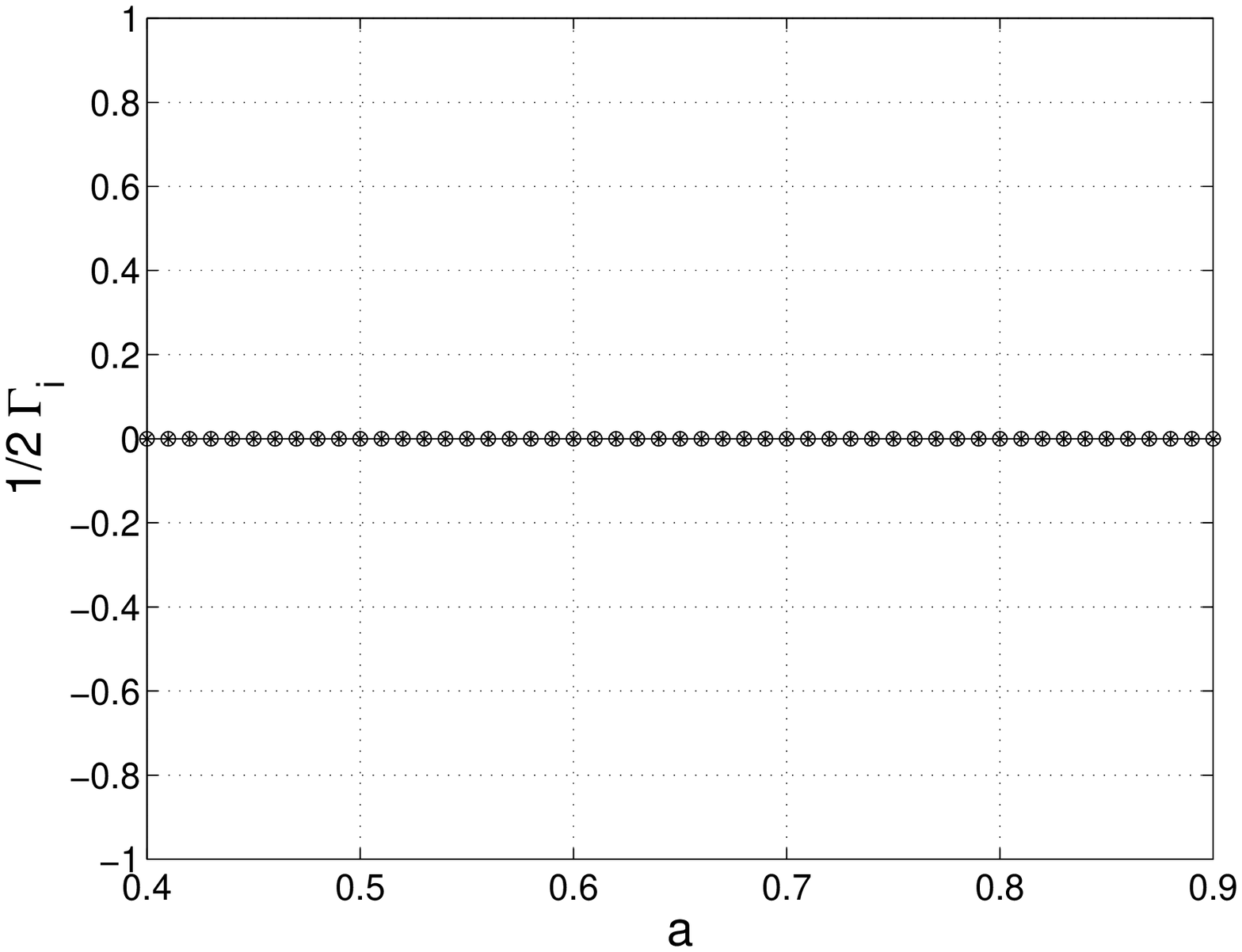,width=7.5cm}
\end{minipage}
\vspace*{.5cm}
\caption{The energies $E_i$ (left) and widths $\Gamma_i /2$
(right) as a function of the
tuning parameter $a$.  $e_1=1-a/2; \; e_2=a $ 
and $\omega = 0.05$. The 
$\gamma_1 /2 $ are 1.10  (top), 
0.90  (middle), 0  (bottom); $\gamma _2 = 1.1 \cdot \gamma _1$. 
The full lines show the $E_i$ and  $\Gamma_i /2$ for $\omega = 0$. 
}
\label{fig:tra1}
\end{figure}

\begin{figure}
\hspace{-1.8cm}
\begin{minipage}[tl]{7.5cm}
\psfig{file=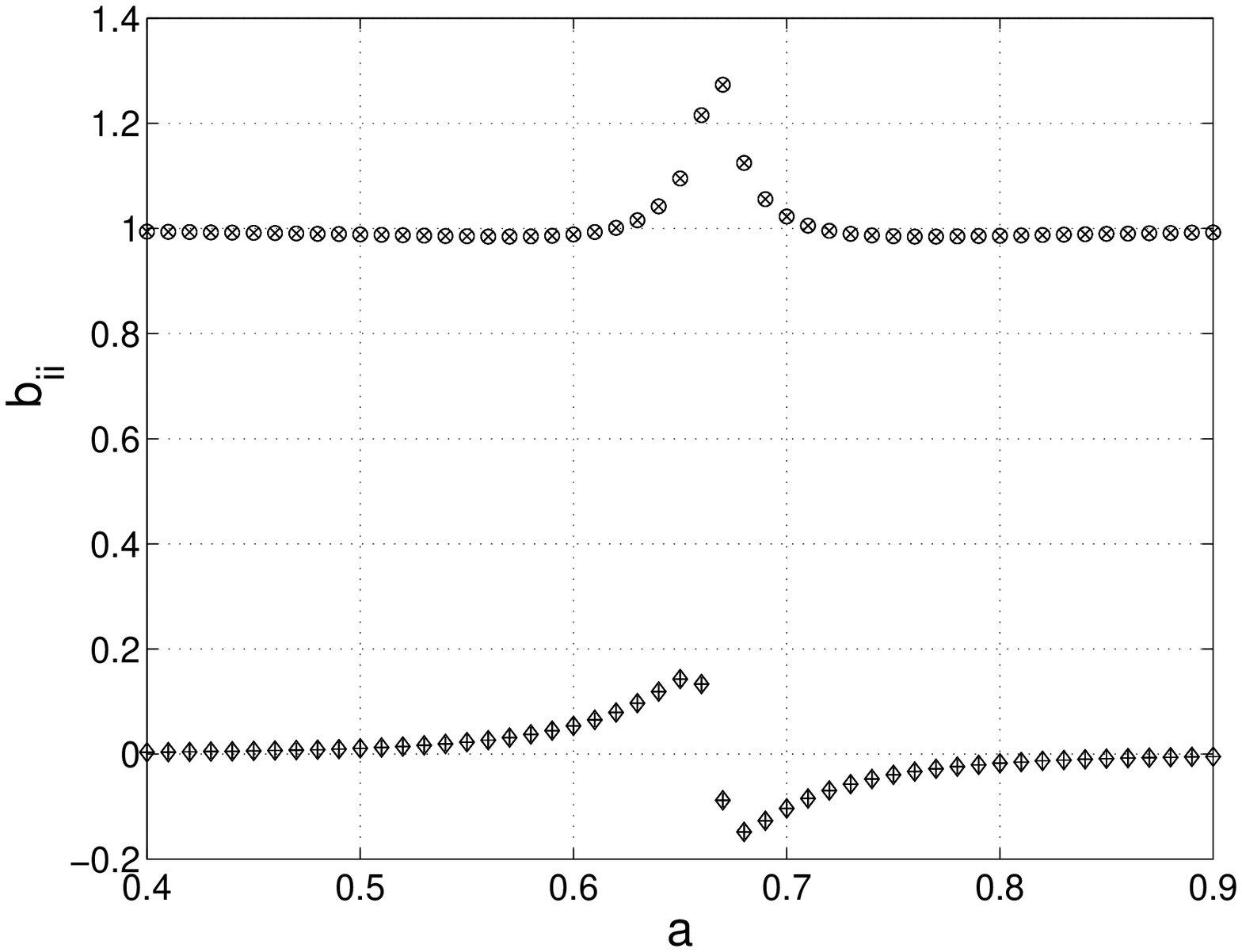,width=7.5cm}
\end{minipage}
\begin{minipage}[tr]{7.5cm}
\psfig{file=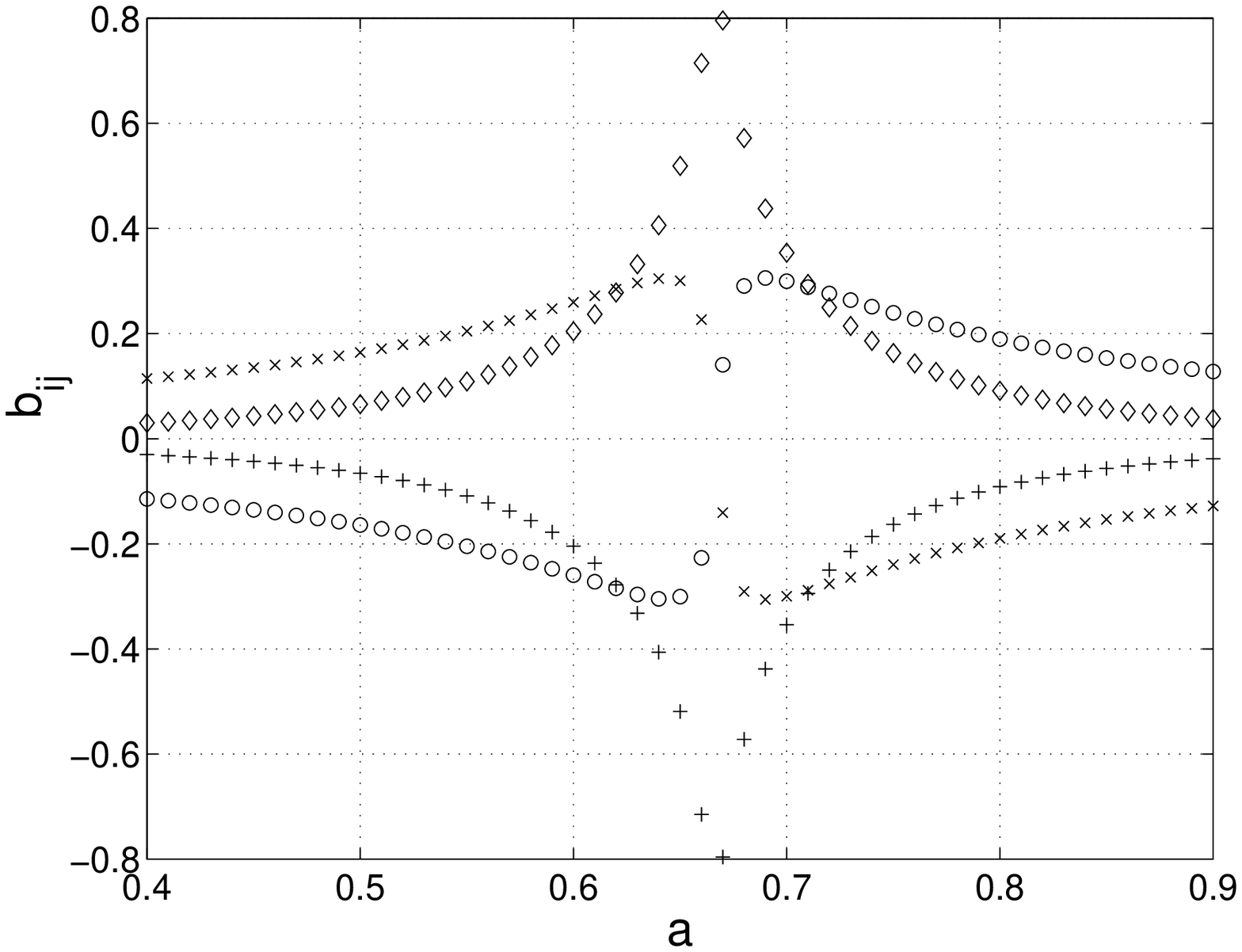,width=7.5cm}
\end{minipage}
\begin{minipage}[ml]{7.5cm}
\psfig{file=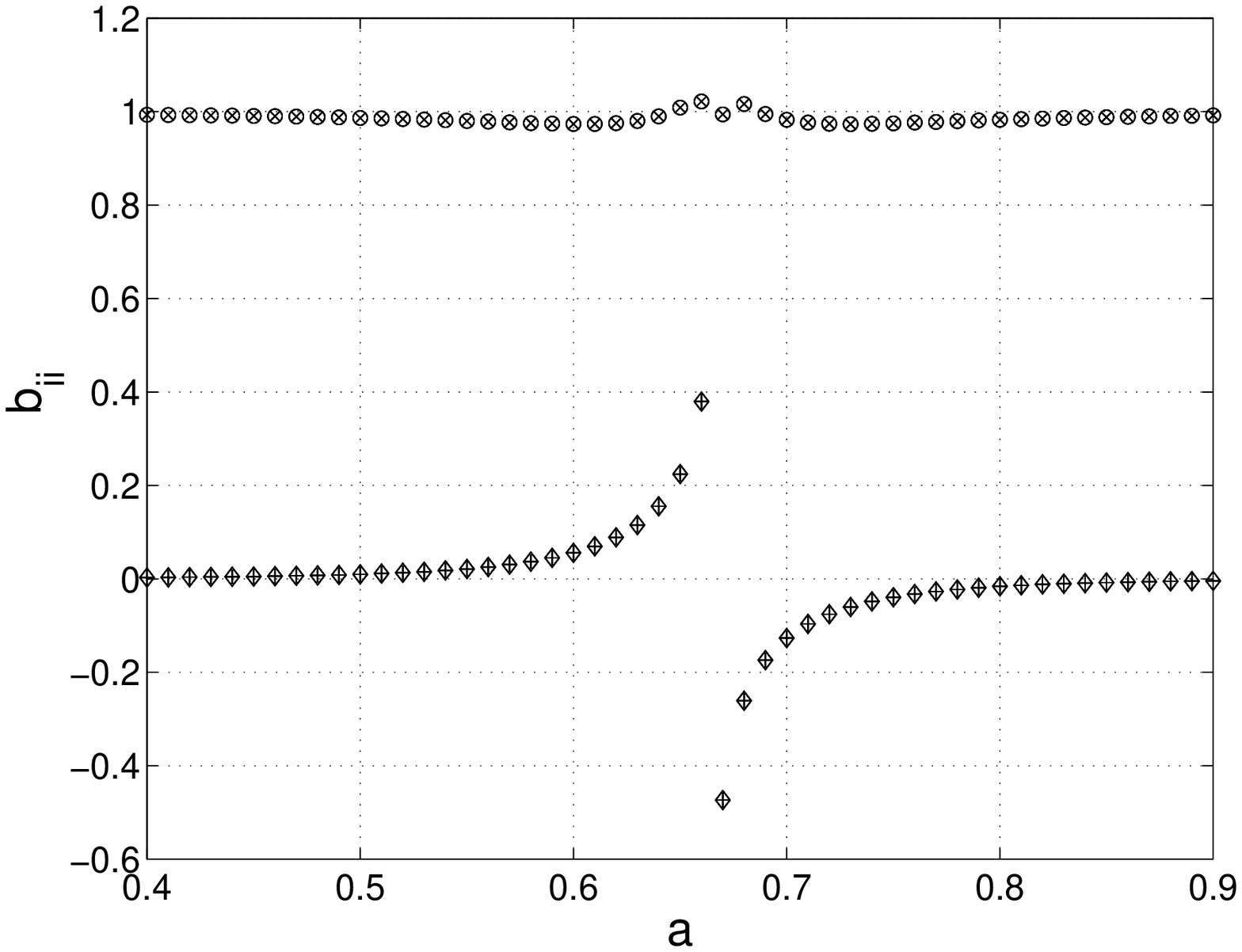,width=7.5cm}
\end{minipage}
\begin{minipage}[mr]{7.5cm}
\psfig{file=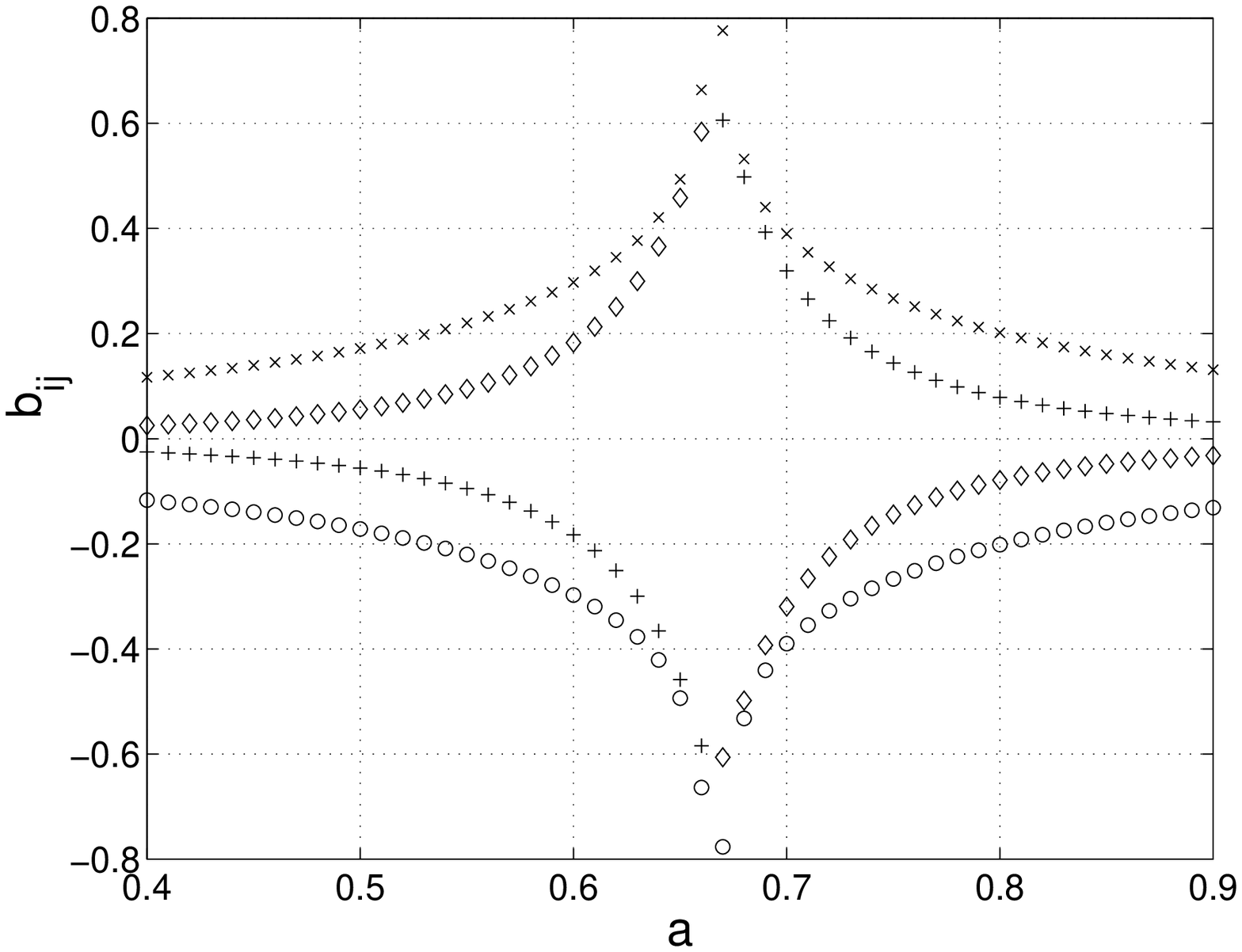,width=7.5cm}
\end{minipage}
\begin{minipage}[bl]{7.5cm}
\psfig{file=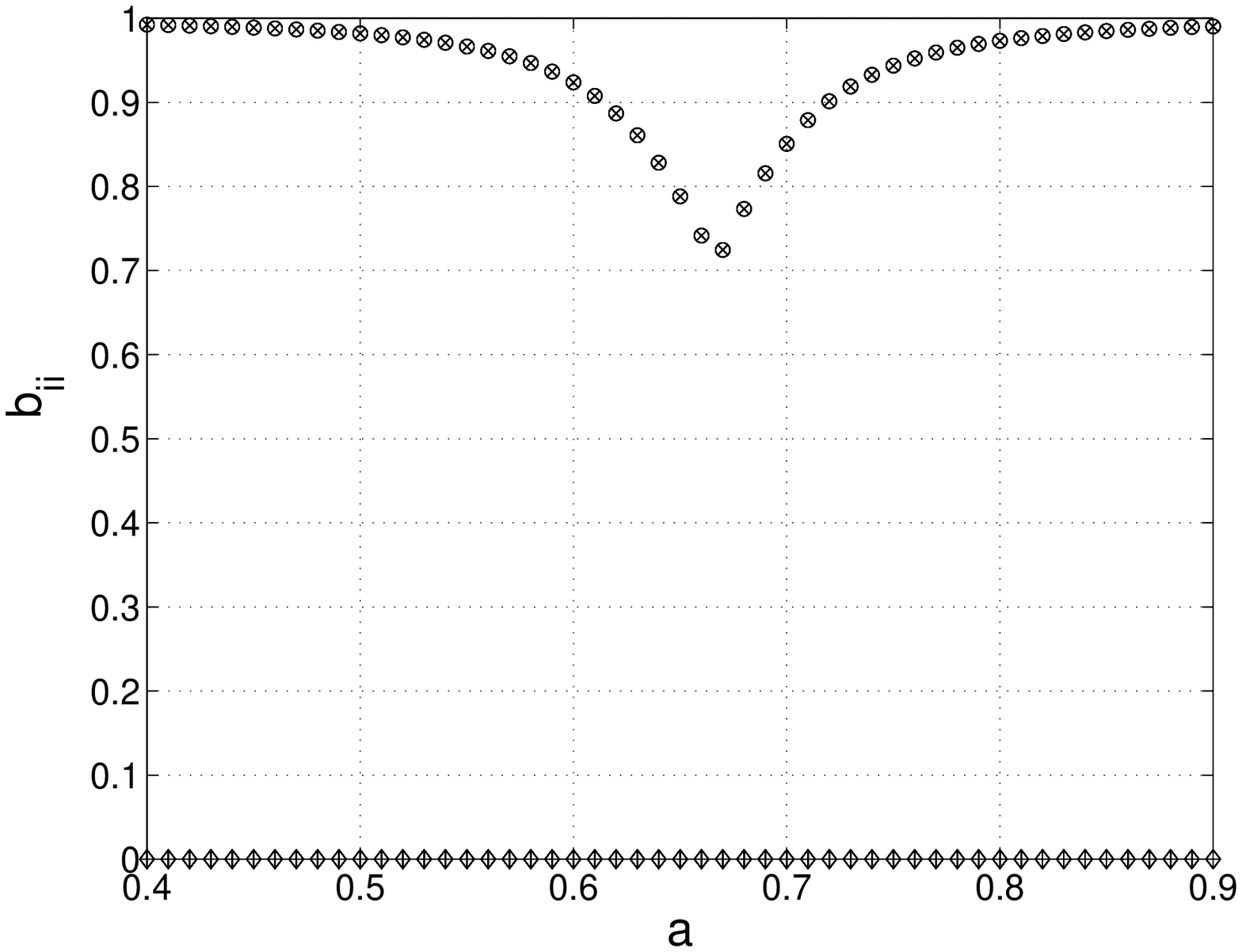,width=7.5cm}
\end{minipage}
\hspace{1.0cm}
\begin{minipage}[br]{7.5cm}
\psfig{file=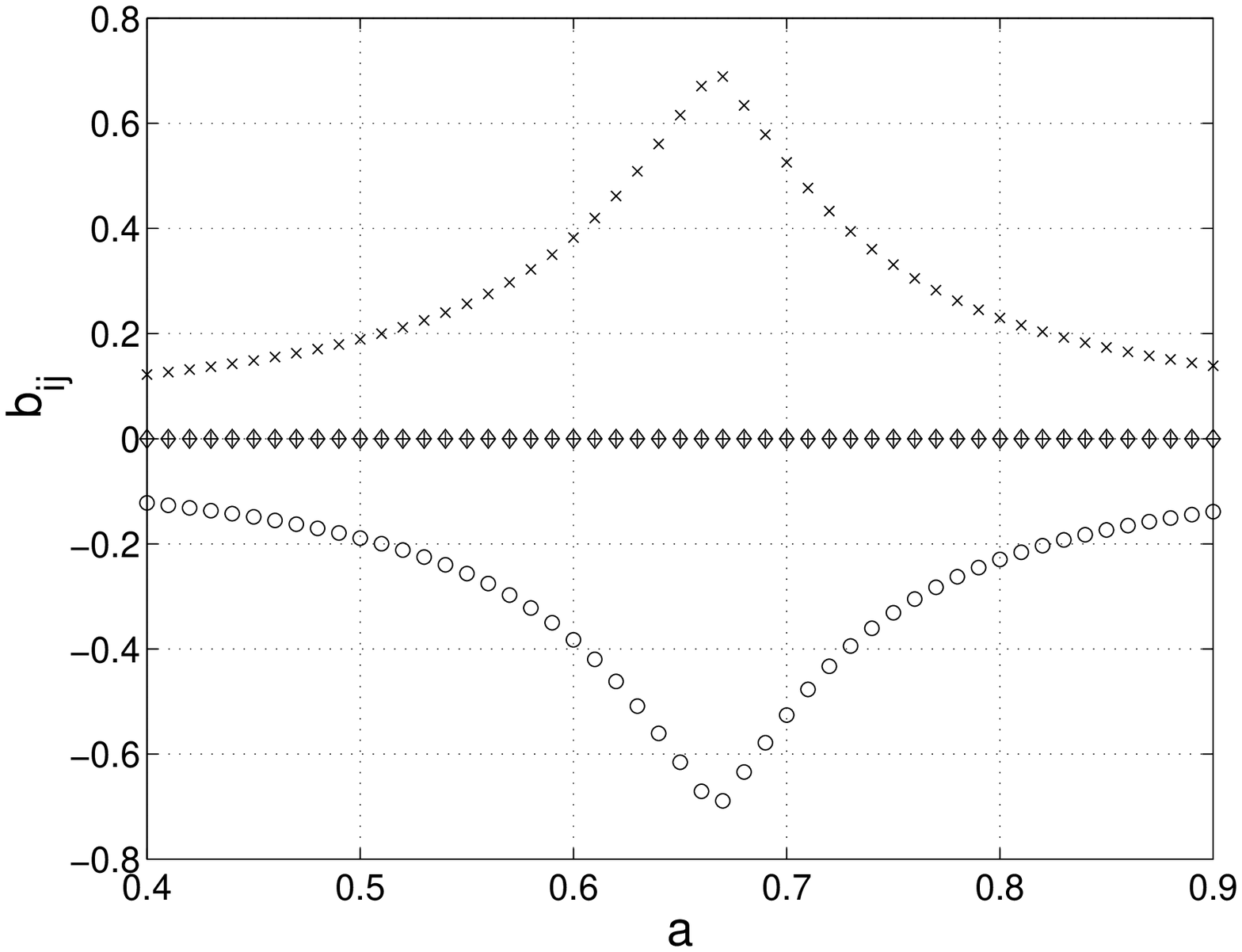,width=7.5cm}
\end{minipage}
\vspace*{.5cm}
\caption{The mixing coefficients $b_{ii}$ (left)
and $b_{ij\ne i}$ (right) defined by
Eq. (\ref{eq:mix}) as a function of the tuning parameter $a$.
o and x denote the real parts and $\diamond$ and + the imaginary parts. 
  $e_1=1-a/2; \; e_2=a $ and $\omega = 0.05$. The 
$\gamma_1 /2 $ are  the same as in Fig. \ref{fig:tra1}:
1.10  (top), 0.90  (middle), 0  (bottom);  $\gamma _2 = 1.1 \cdot \gamma _1$.
Note the different scales in the three cases.
}
\label{fig:tra3}
\end{figure}

\begin{figure}
\begin{minipage}[tl]{7.5cm}
\psfig{file=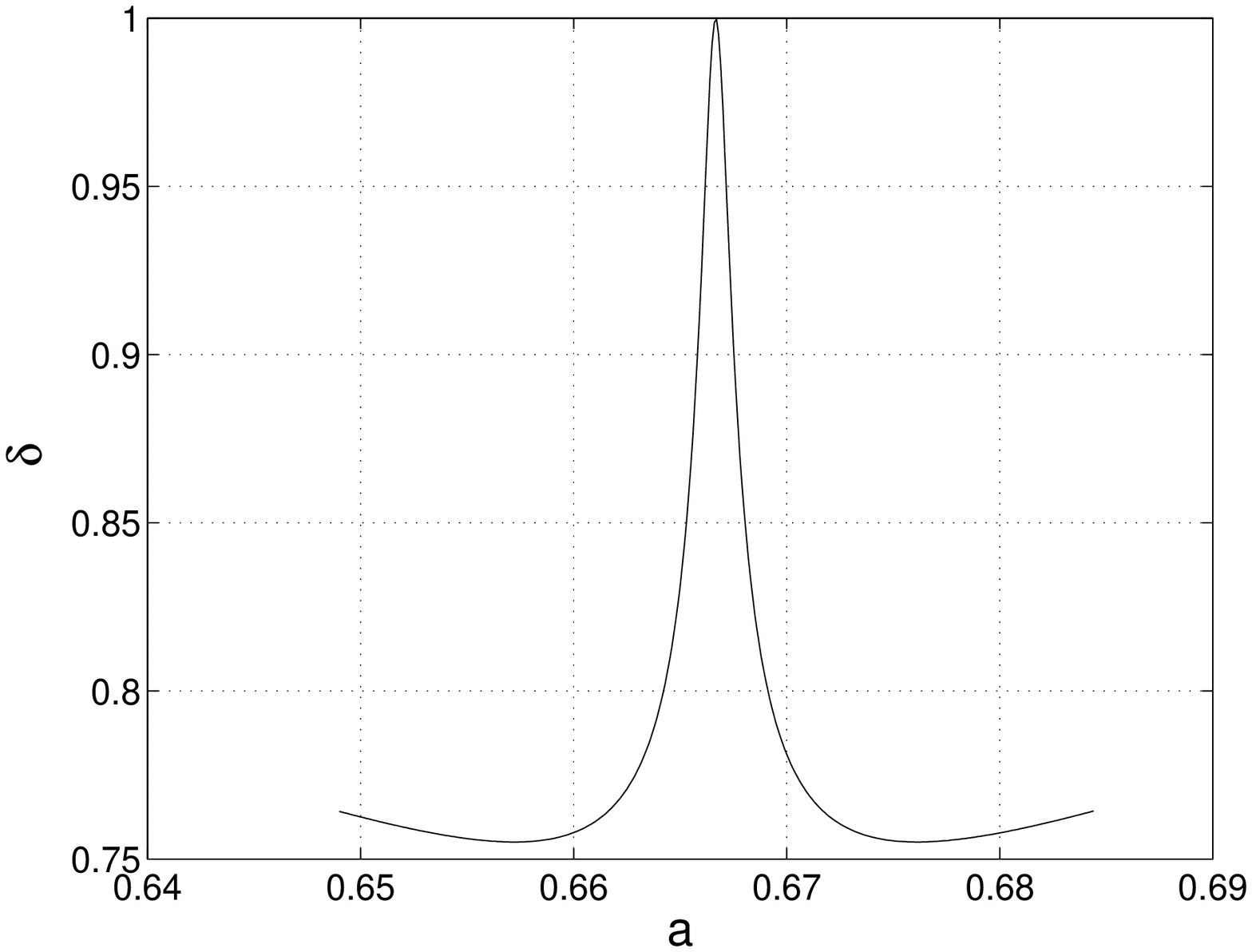,width=7.5cm}
\end{minipage}
\begin{minipage}[tr]{7.5cm}
\psfig{file=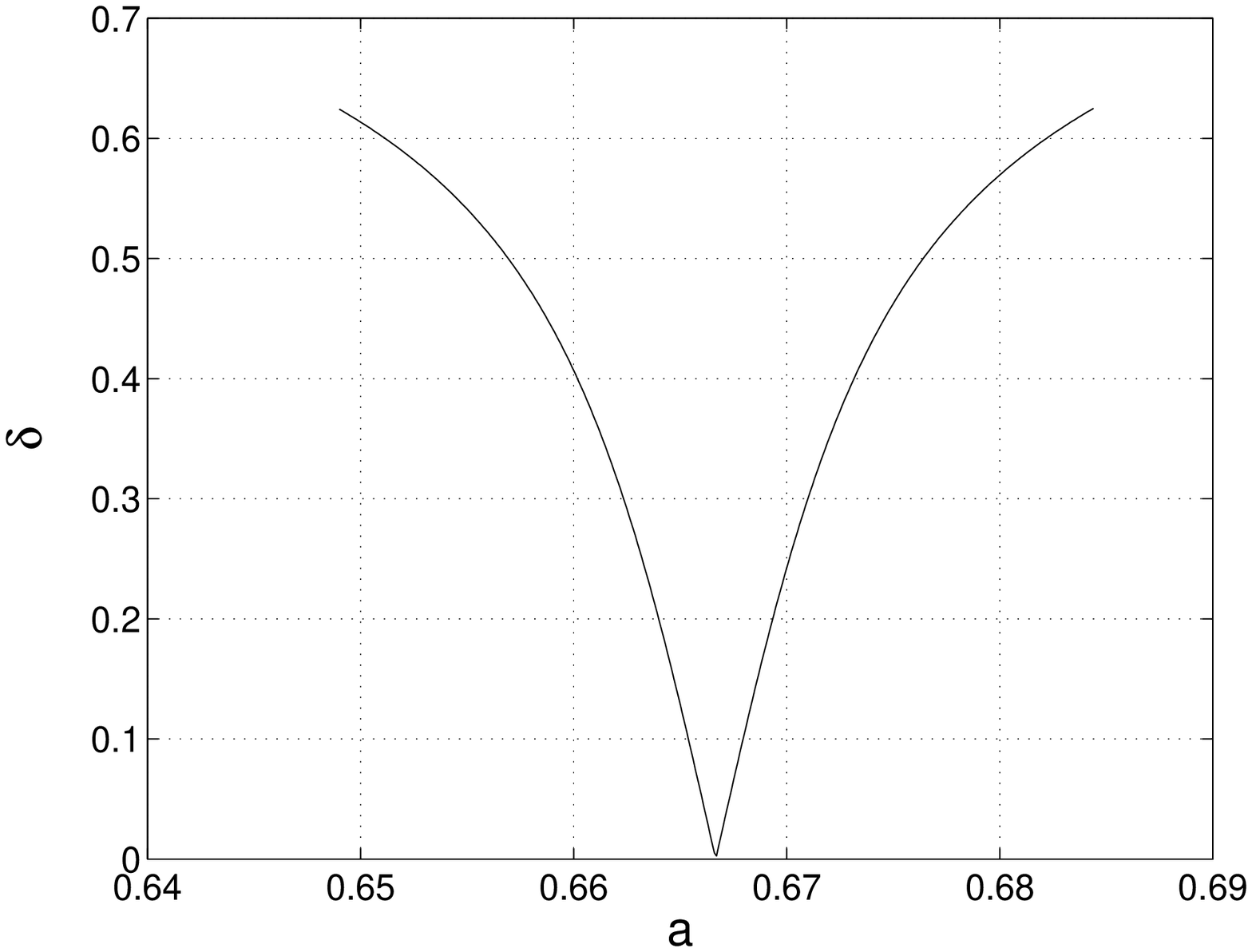,width=7.5cm}
\end{minipage}
\hspace*{.5cm}
\begin{minipage}[bl]{7.5cm}
\psfig{file=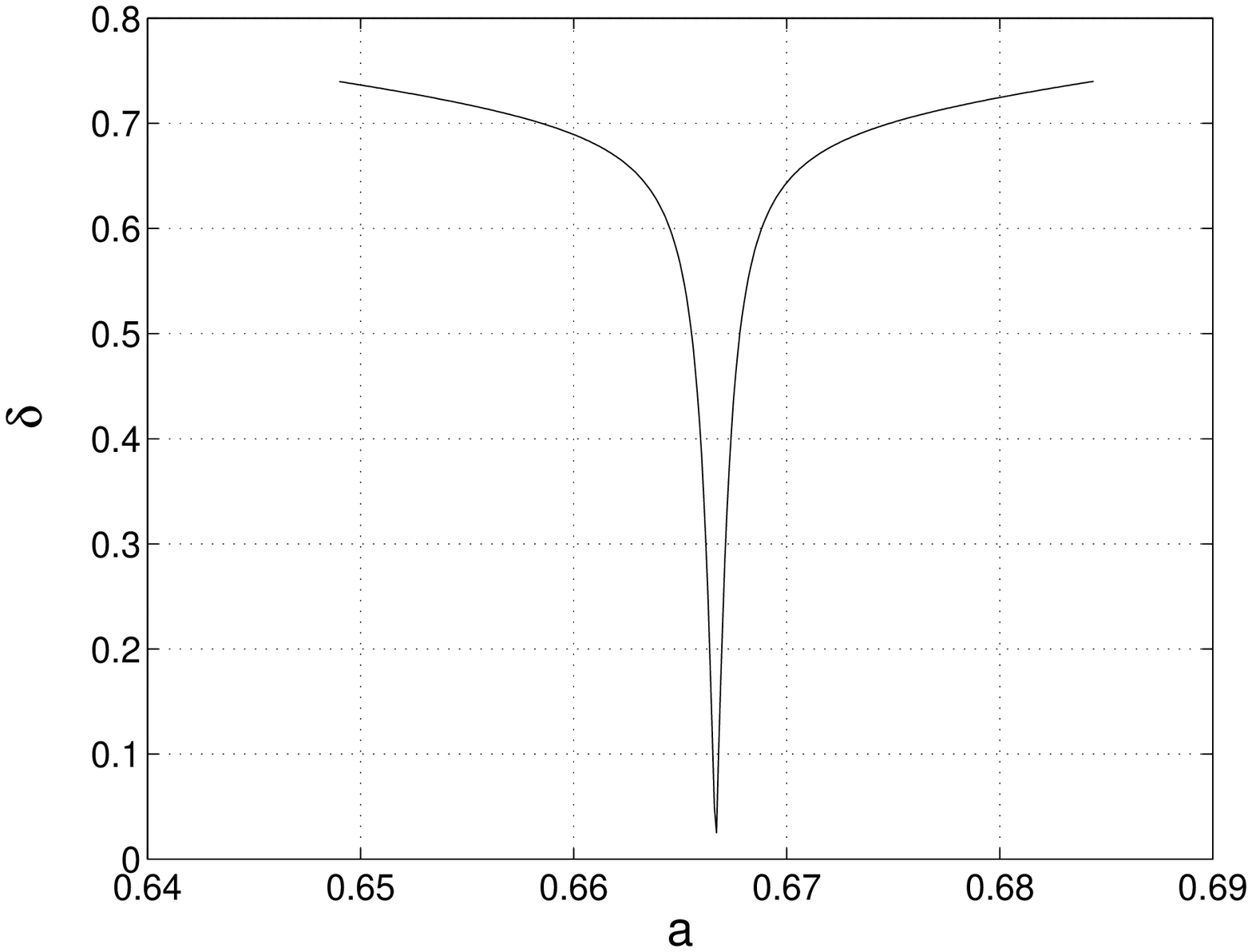,width=7.5cm}
\end{minipage}
\hspace*{.6cm}
\begin{minipage}[br]{7.5cm}
\psfig{file=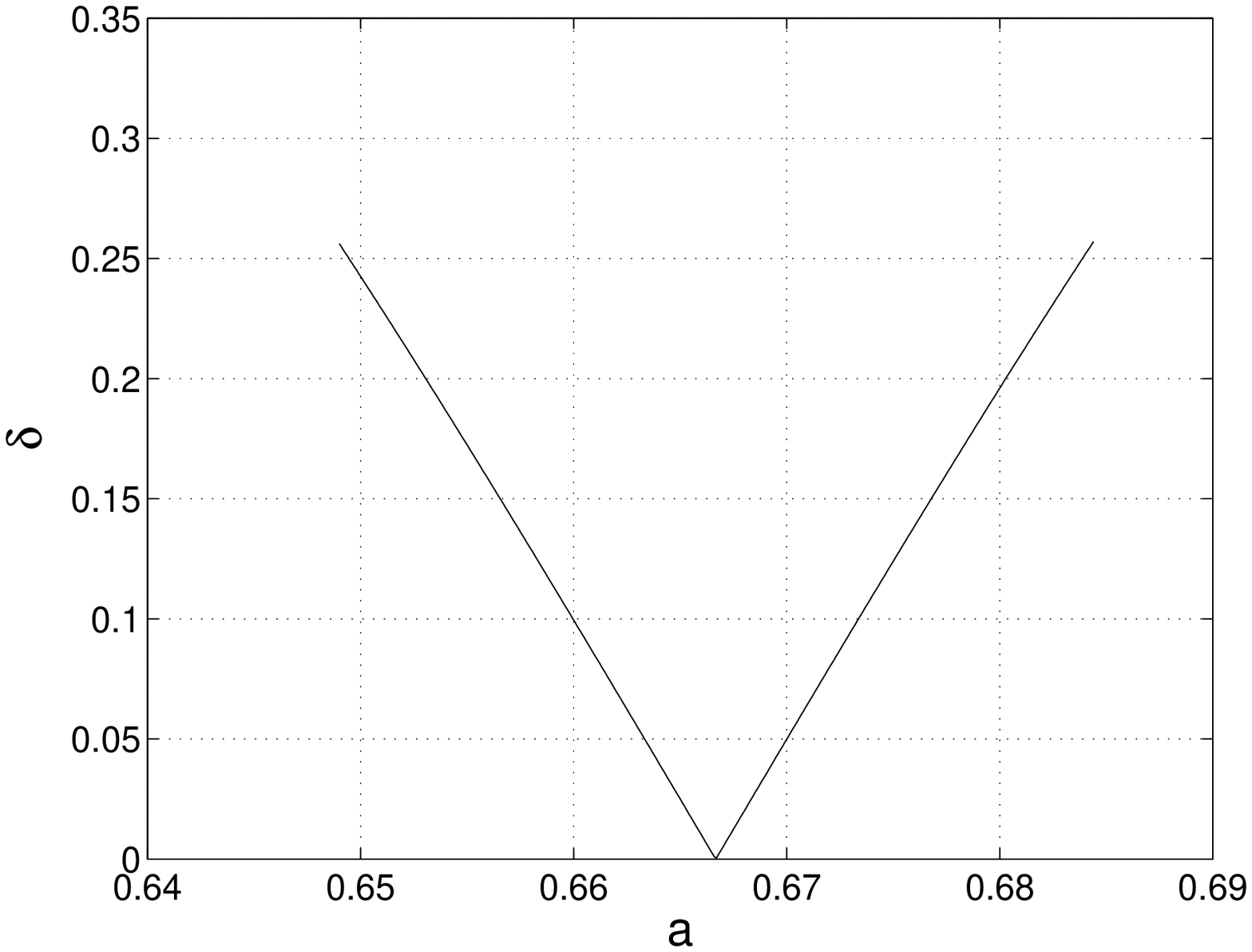,width=7.5cm}
\end{minipage}
\vspace*{.5cm}
\caption{The differences  $\delta= |b_{ii}|^2 - |b_{ij\ne i}|^2$ 
 as a function of the tuning parameter $a$.
 $e_1=1-a/2; \; e_2=a $ and $\omega = 0.05$. The 
$\gamma_1 /2 $  are 1.010  (top left), 0.990  (bottom left), 
0.90  (top right), 0  (bottom right); $\gamma _2 = 1.1 \cdot \gamma _1$.
Note the different scales in the  different figures.
}
\label{fig:avoi1}
\end{figure}

\begin{figure}
\begin{minipage}[tl]{7.5cm}
\psfig{file=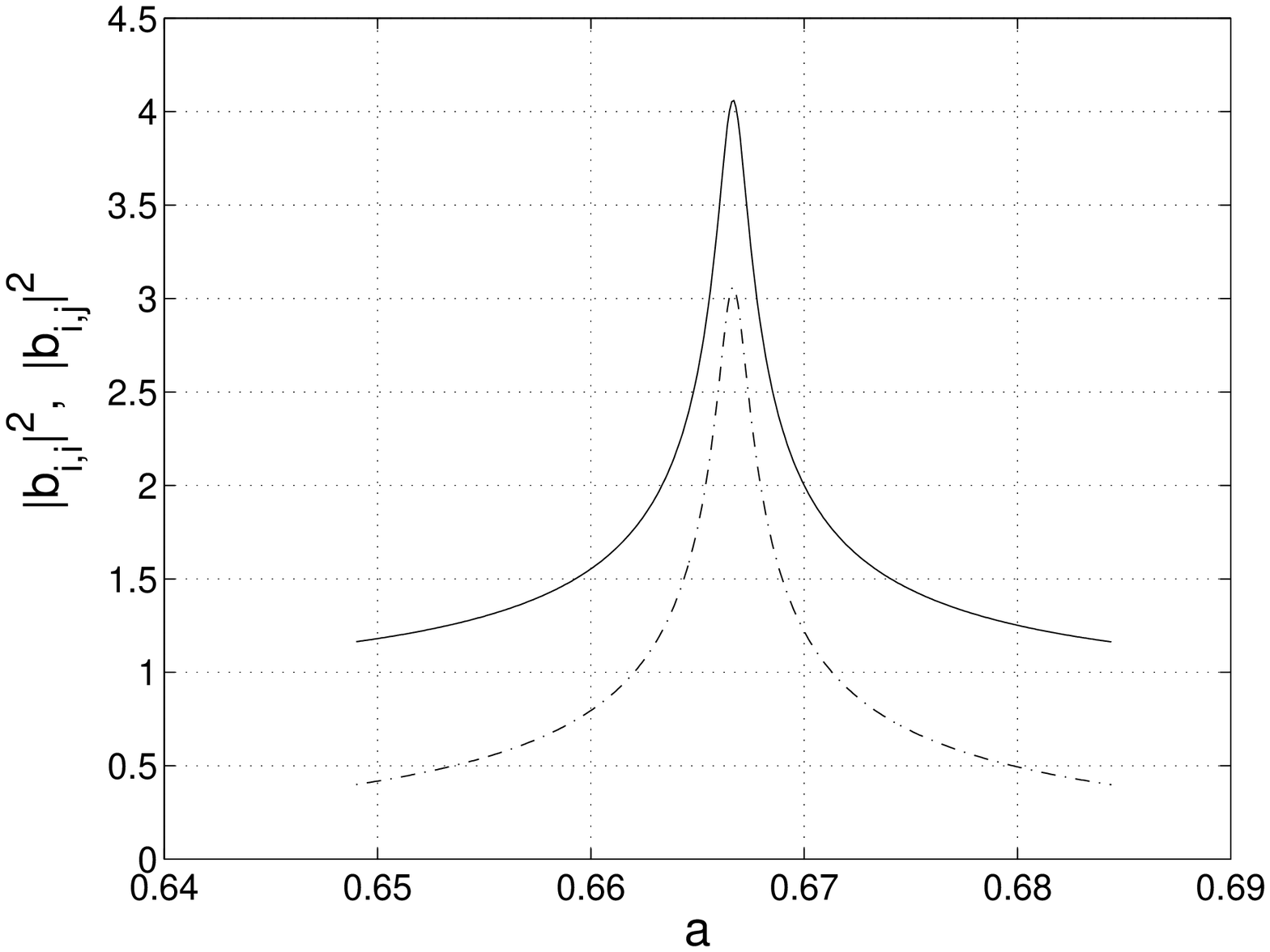,width=7.5cm}
\end{minipage}
\begin{minipage}[tr]{7.5cm}
\psfig{file=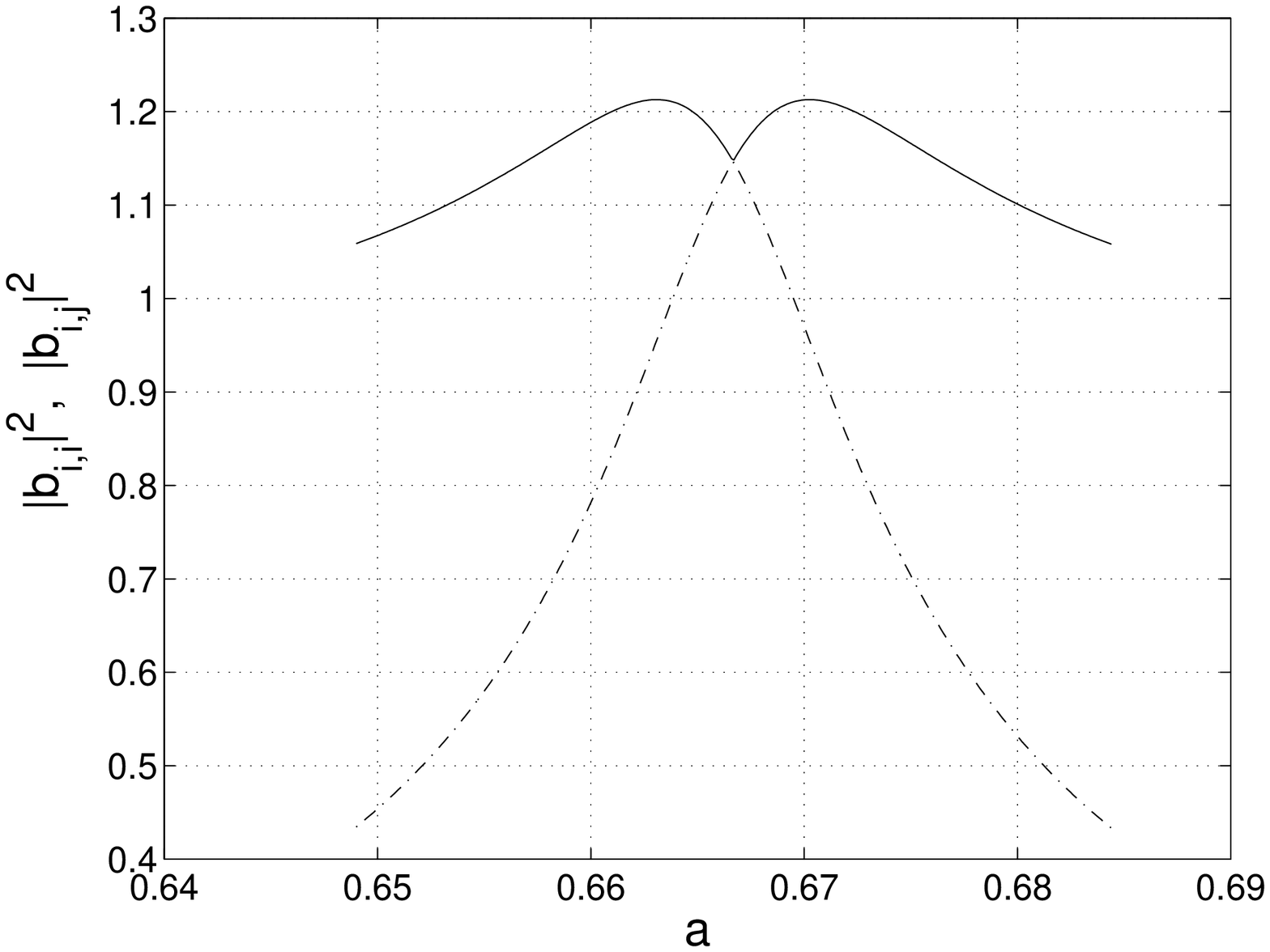,width=7.5cm}
\end{minipage}
\hspace*{.5cm}
\begin{minipage}[bl]{7.5cm}
\psfig{file=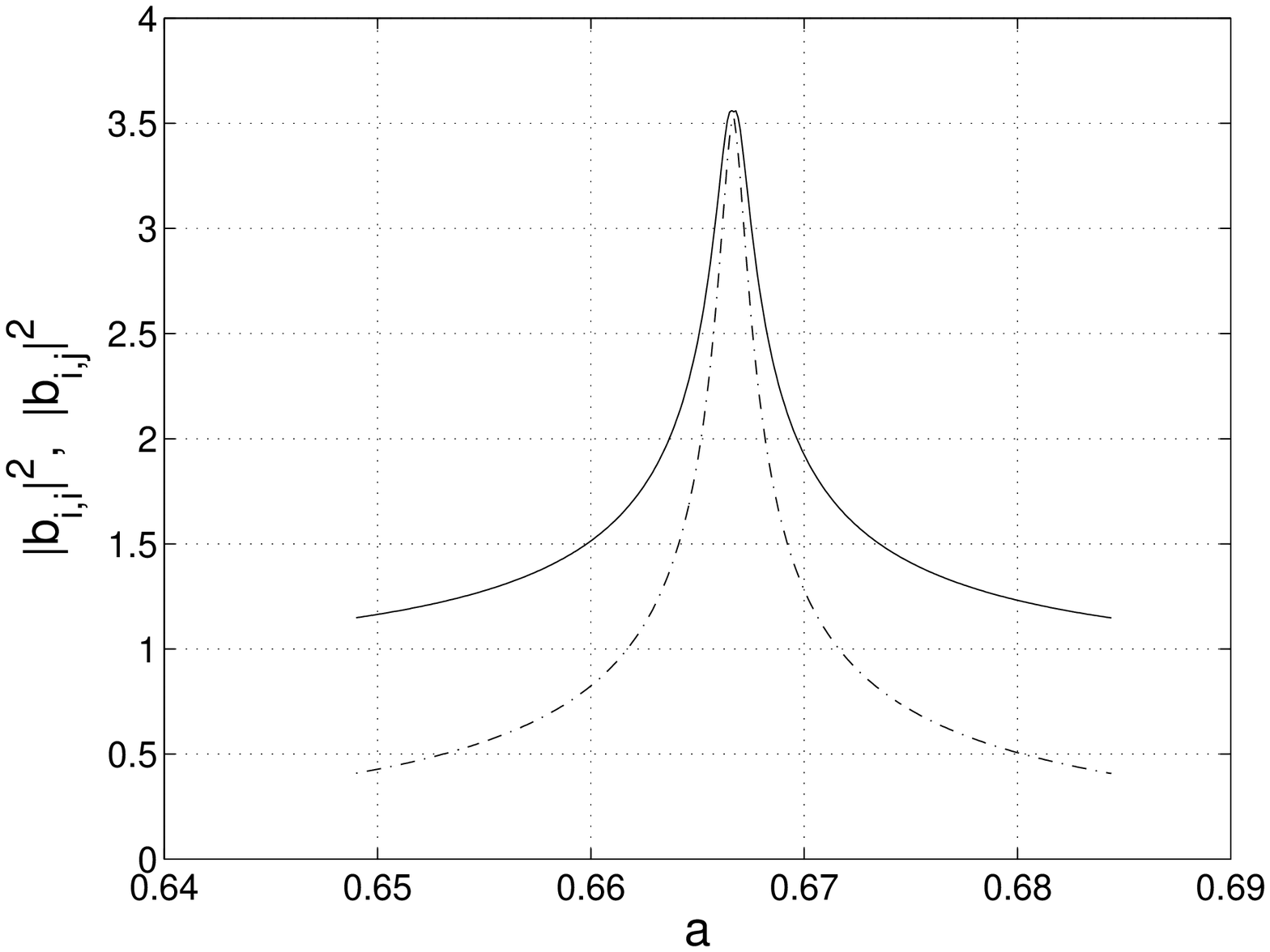,width=7.5cm}
\end{minipage}
\hspace*{.6cm}
\begin{minipage}[br]{7.5cm}
\psfig{file=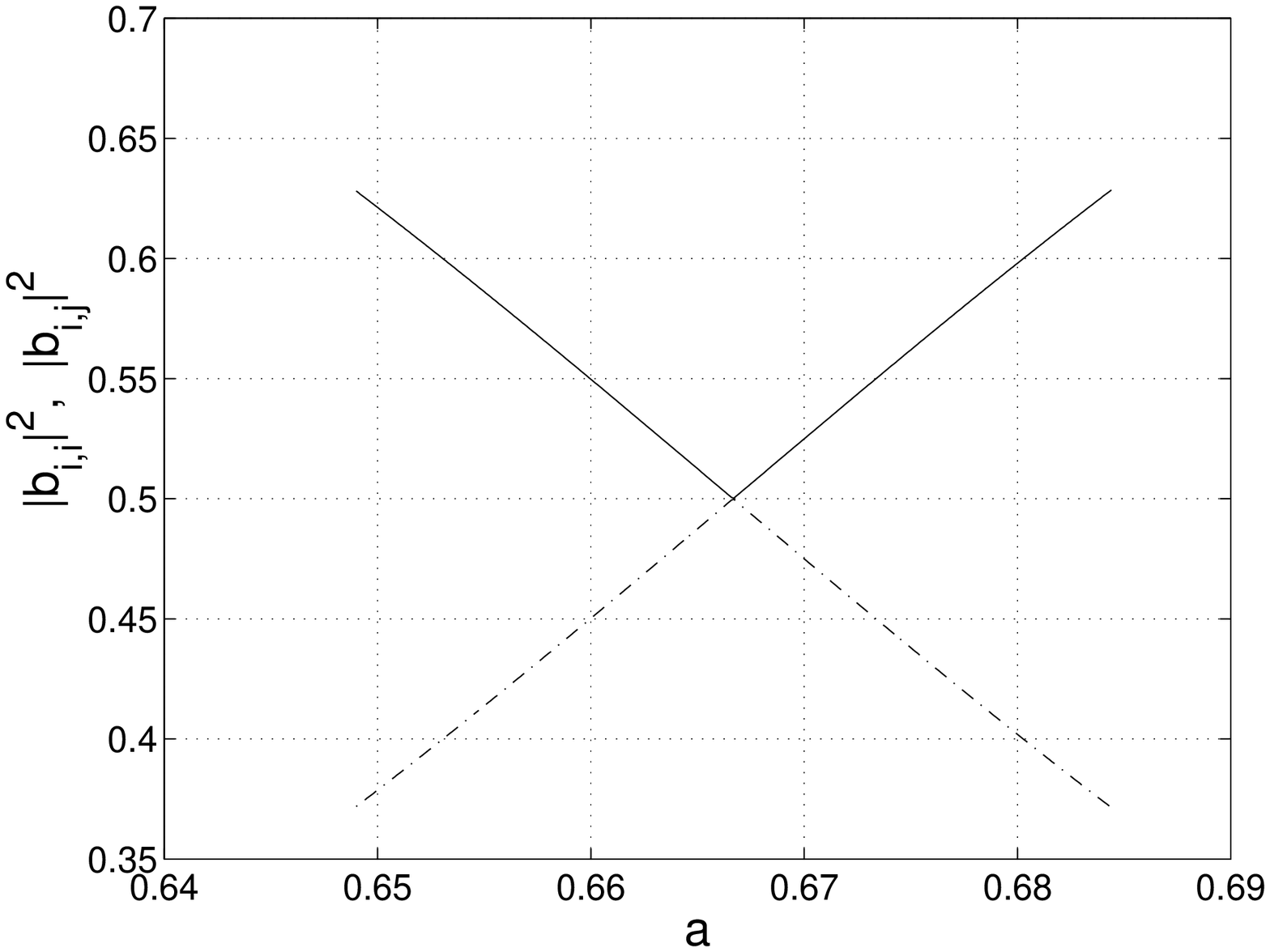,width=7.5cm}
\end{minipage}
\vspace*{.5cm}
\caption{
The  $|b_{ii}|^2$ (full lines)  and $ |b_{ij\ne i}|^2$ (dash-dotted lines) 
 as a function of the tuning parameter $a$.
 $e_1=1-a/2; \; e_2=a $ and $\omega = 0.05$. The $\gamma_1 /2 $ are 
the same  as in 
Fig. \ref{fig:avoi1}: 1.010  (top left), 0.990  (bottom left), 
0.90  (top right), 0  (bottom right); $\gamma _2 = 1.1 \cdot \gamma _1$.
Note the different scales in the  different figures.
}
\label{fig:avoi2}
\end{figure}

The exchange of the wave functions continues analytically into the
function space of discrete states as illustrated in Figs.  \ref{fig:tra3},
 \ref{fig:avoi1}   and  \ref{fig:avoi2}.
When the resonance states avoid  crossing at  $a^{\rm cr}$, 
the components of the wave functions do not increase up to infinity.
Their increase is reduced due to interferences (Fig. \ref{fig:avoi2}).
 The differences $\delta = |b_{ii}|^2 - |b_{ij\ne i}|^2$  jump 
from 1 to 0 at $a^{\rm cr}$ (and from about 0 to almost 1 
for  values of $ a$ distant from $a^{\rm cr}$)   when
$ (\gamma_i - \gamma_i^{\rm cr})$ changes its sign
 (Fig. \ref{fig:avoi1}). This jump 
is related to the exchange of the wave functions.
The value  $\delta = 0$ at $a^{\rm cr}$ remains unaltered
when $(\gamma_1 /2 - \gamma_2 /2)^2
 < 4 \; \omega^2$, i.e. also for discrete states.
 Therefore, the results shown
in Fig. \ref{fig:tra1} top and middle correspond to situations being 
fundamentally
and topologically different from one another.

\begin{figure}
\begin{minipage}[tl]{7.5cm}
\psfig{file=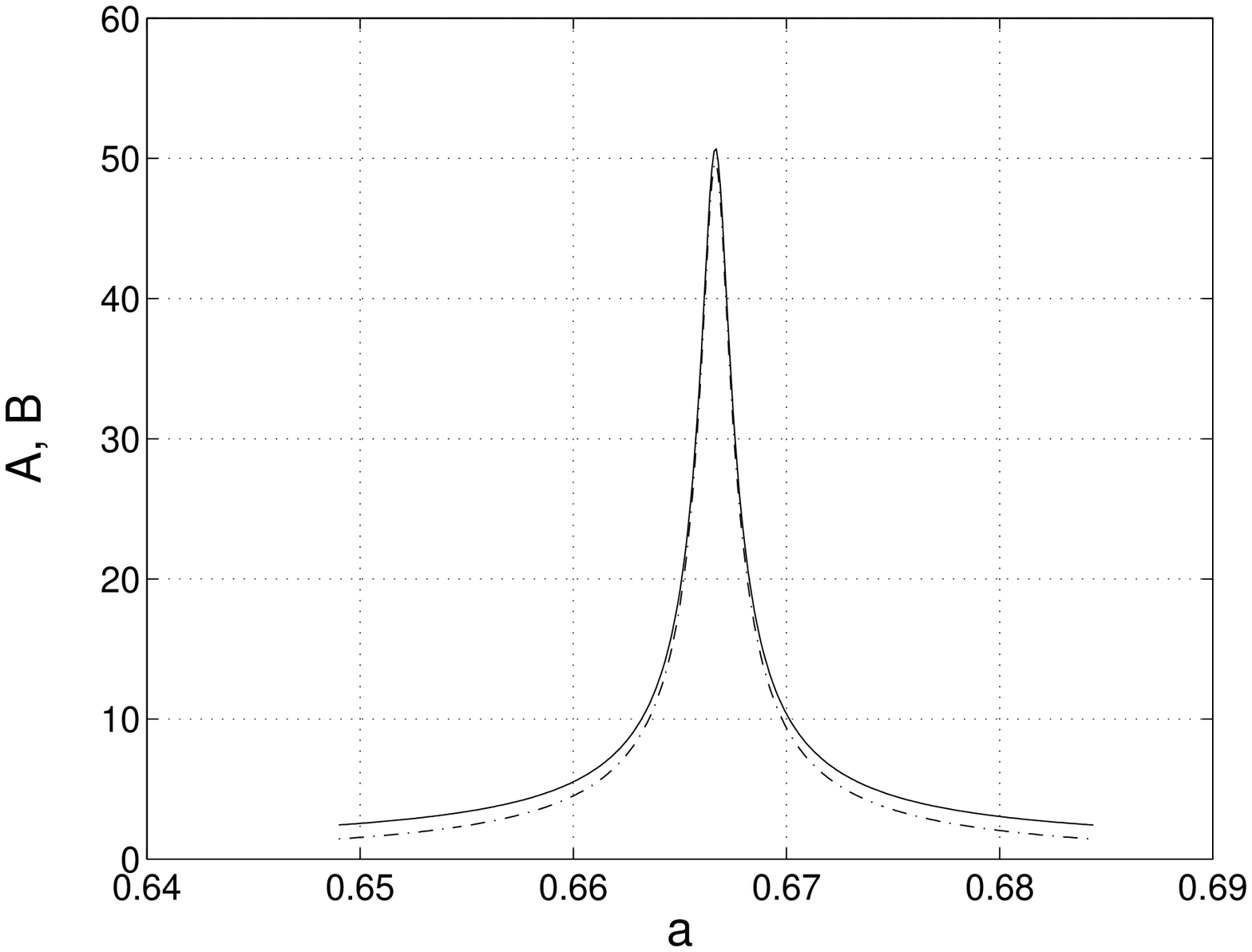,width=7.5cm}
\end{minipage}
\begin{minipage}[tr]{7.5cm}
\psfig{file=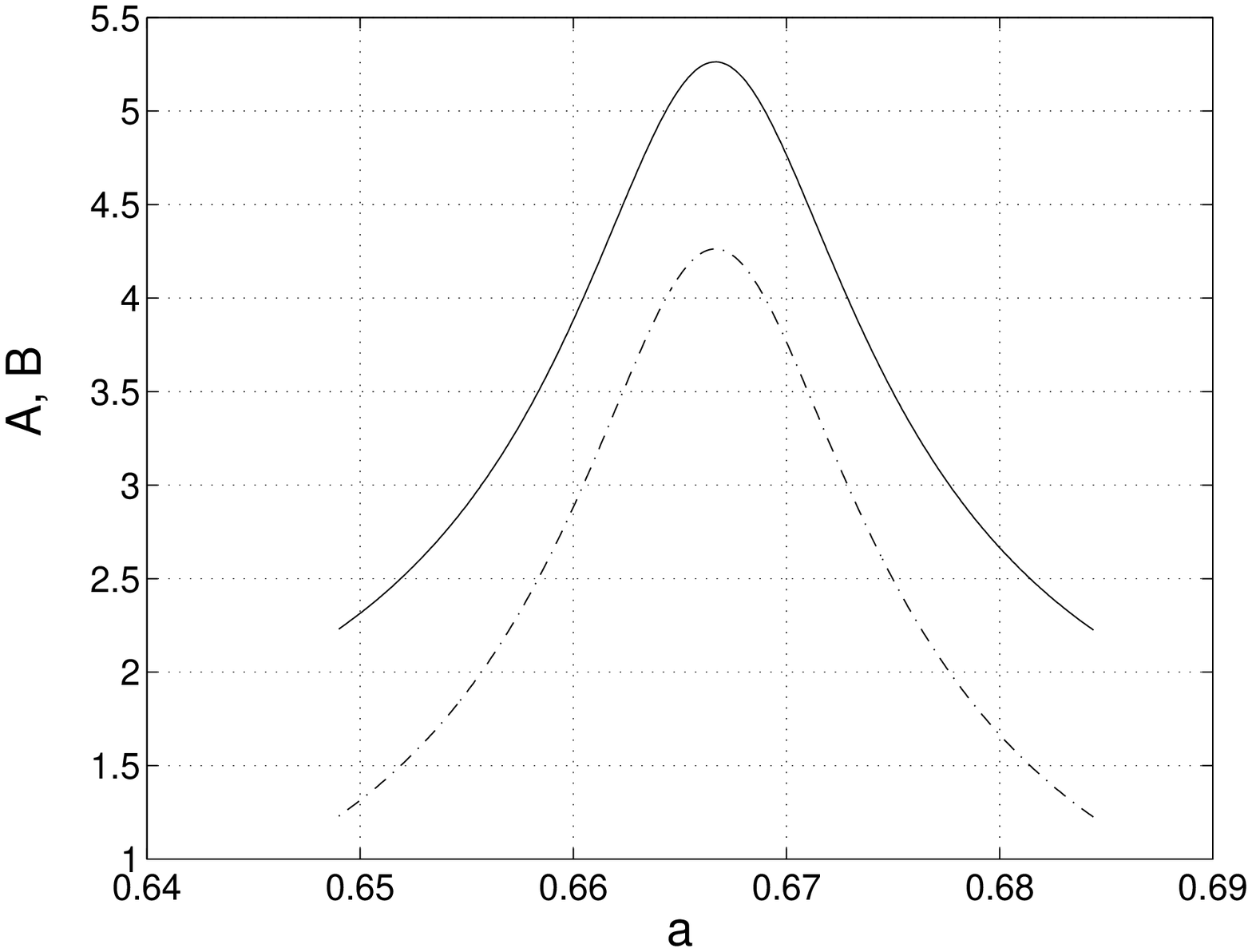,width=7.5cm}
\end{minipage}
\hspace*{.5cm}
\begin{minipage}[bl]{7.5cm}
\psfig{file=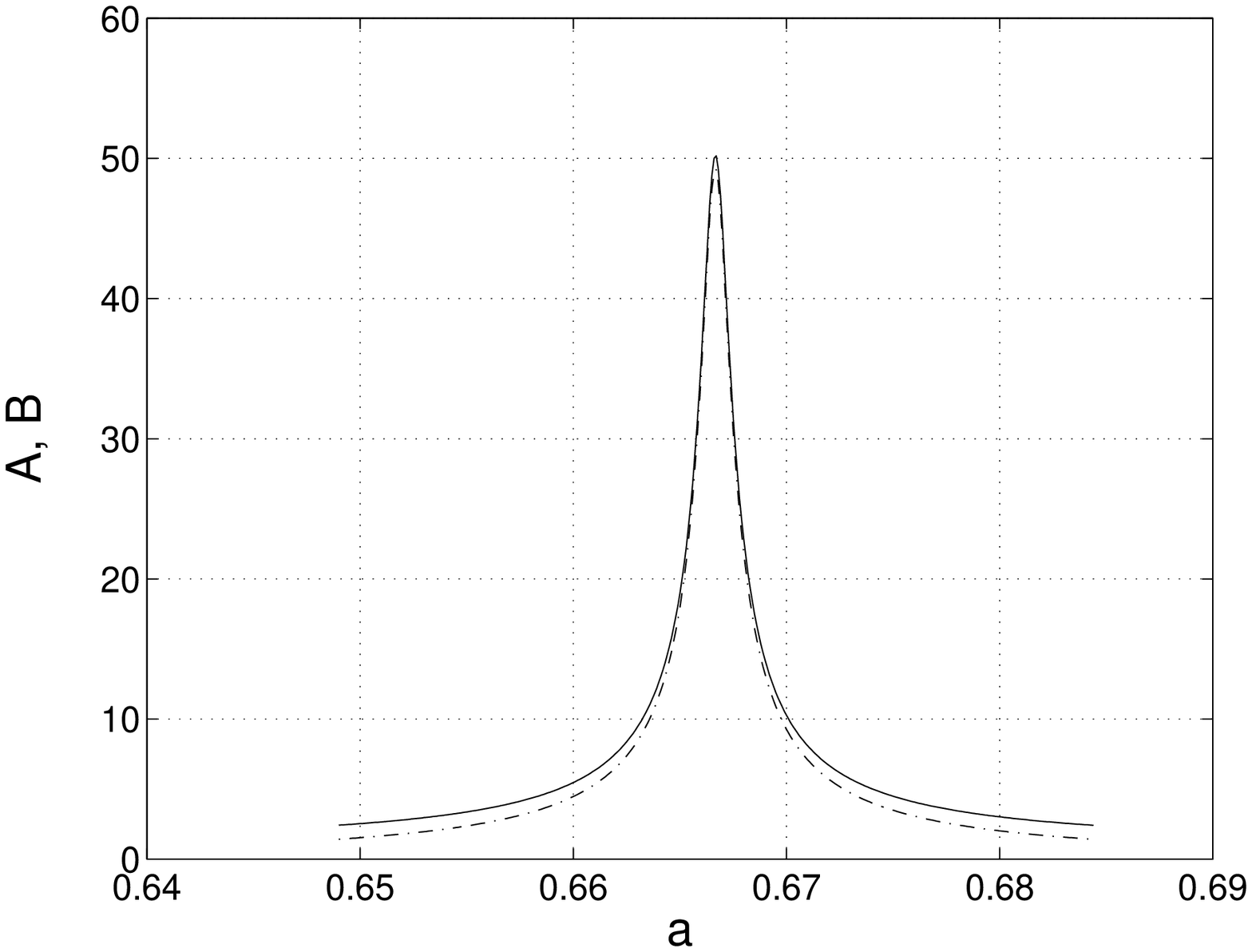,width=7.5cm}
\end{minipage}
\hspace*{.6cm}
\begin{minipage}[br]{7.5cm}
\psfig{file=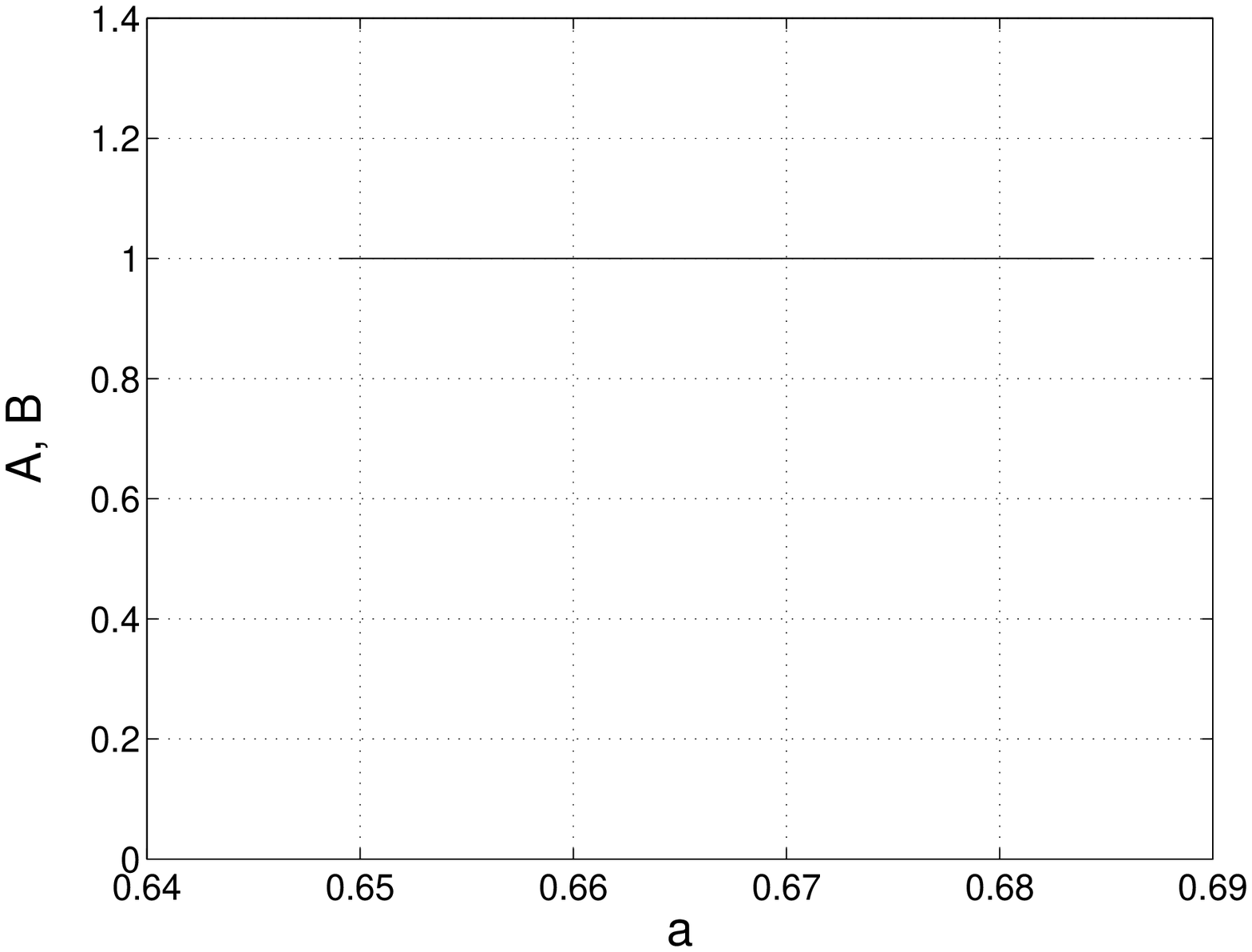,width=7.5cm}
\end{minipage}
\vspace*{.5cm}
\caption{
The  $A$ (full lines)  and $ B$ (dash-dotted lines) defined in Eq.
(\ref{eq:biorth})
 as a function of the tuning parameter $a$.
  $e_1=1-a/2; \; e_2=a $ and $\omega = 0.05$. The 
$\gamma_1 /2 $  are the same as in 
Fig. \ref{fig:avoi1}:  1.010  (top left), 0.990  (bottom left), 
0.90  (top right), 0  (bottom right); $\gamma _2 = 1.1 \cdot \gamma _1$.
Note the different scales in the  different figures.
}
\label{fig:avoi3}
\end{figure}

The bi-orthogonality (\ref{eq:biorth}) of the wave functions 
is characteristic of the
avoided crossing of resonance states. It increases limitless at the double
pole of the $S$ matrix and vanishes in the case of an 
avoided crossing of discrete states (Fig. \ref{fig:avoi3}).
It does not enter  any physically relevant values since it 
does not enter  the $S$ matrix, Eq. (\ref{eq:smatres}). 
The wave functions of the resonance states 
appear in the $S$ matrix in accordance with the 
orthogonality relations (\ref{eq:orth}) which are fulfilled 
in the whole function space without any exceptions.

\begin{figure}
\hspace{-1.8cm}
\begin{minipage}[tl]{7.5cm}
\psfig{file=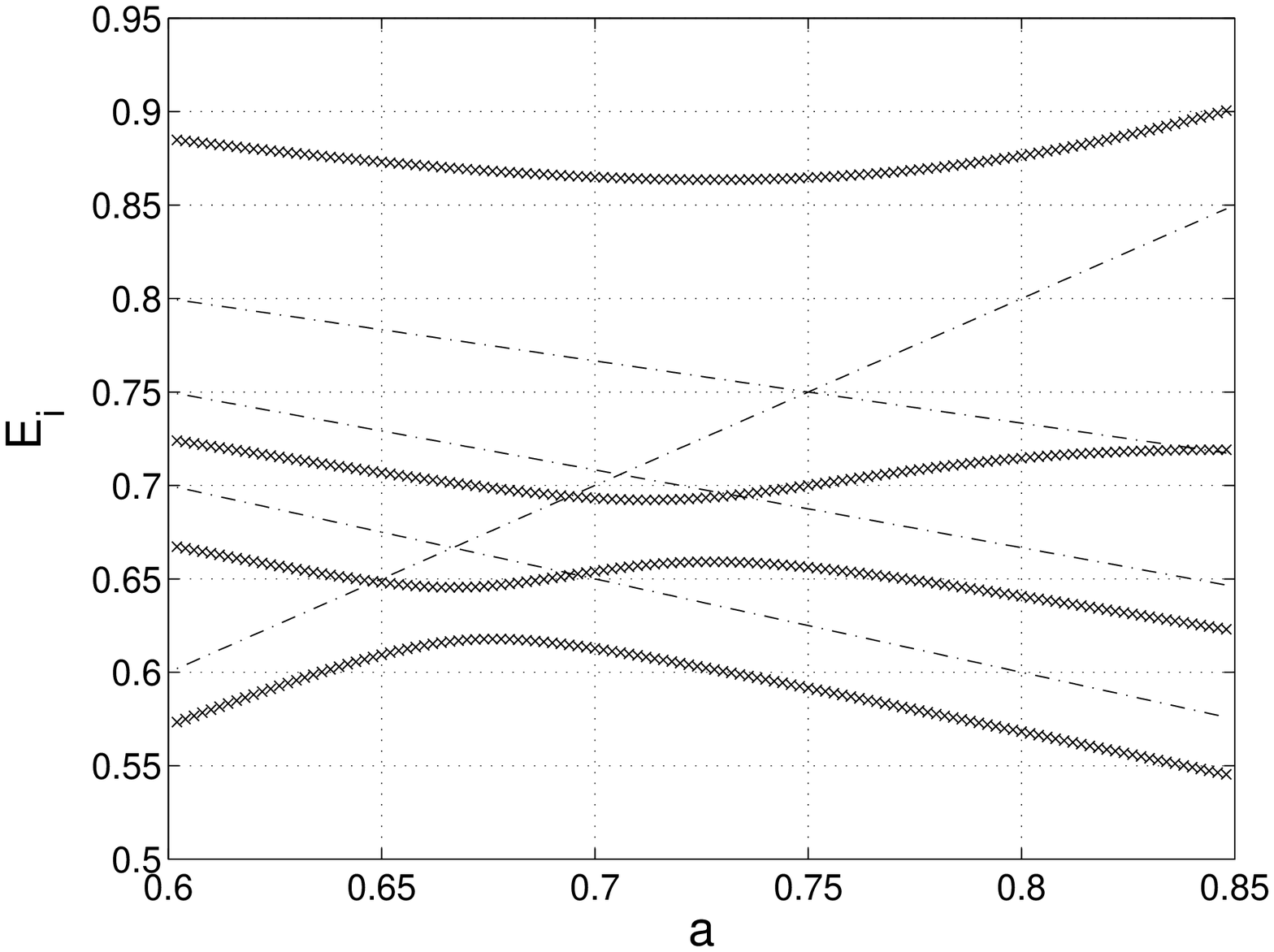,width=7.5cm}
\end{minipage}
\begin{minipage}[tr]{7.5cm}
\psfig{file=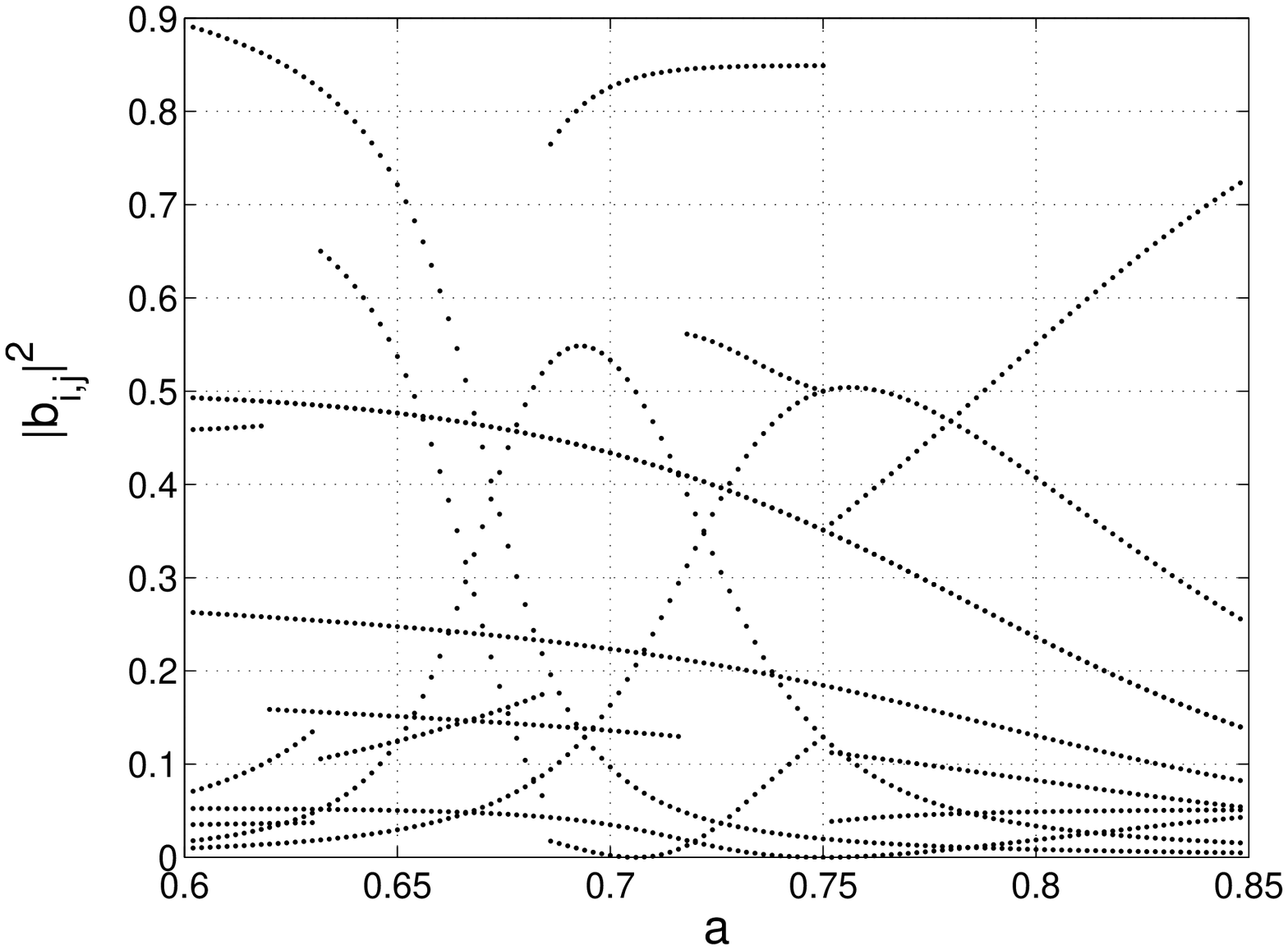,width=7.5cm}
\end{minipage}
\hspace{1cm}
\begin{minipage}[ml]{7.5cm}
\psfig{file=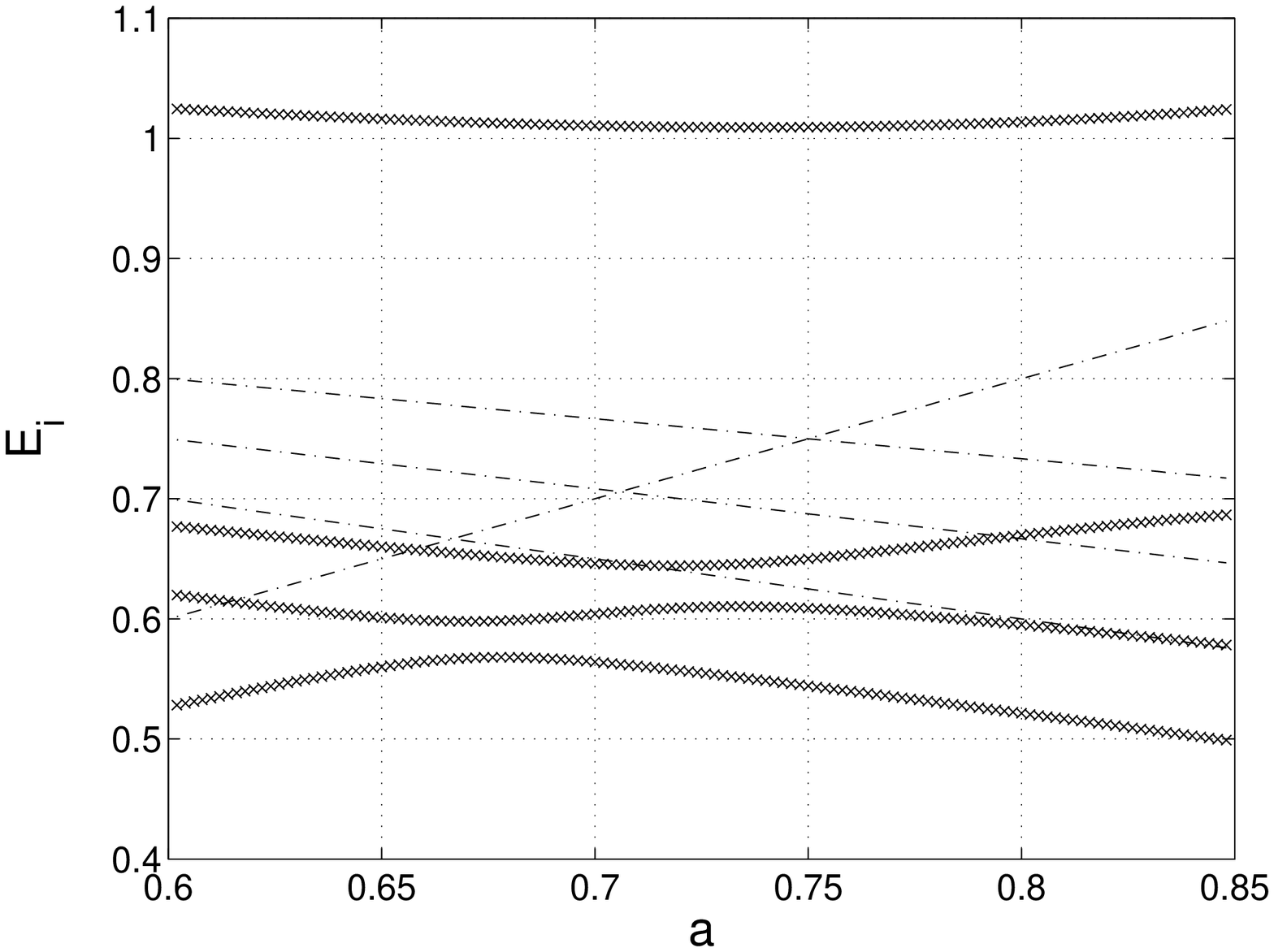,width=7.5cm}
\end{minipage}
\begin{minipage}[mr]{7.5cm}
\psfig{file=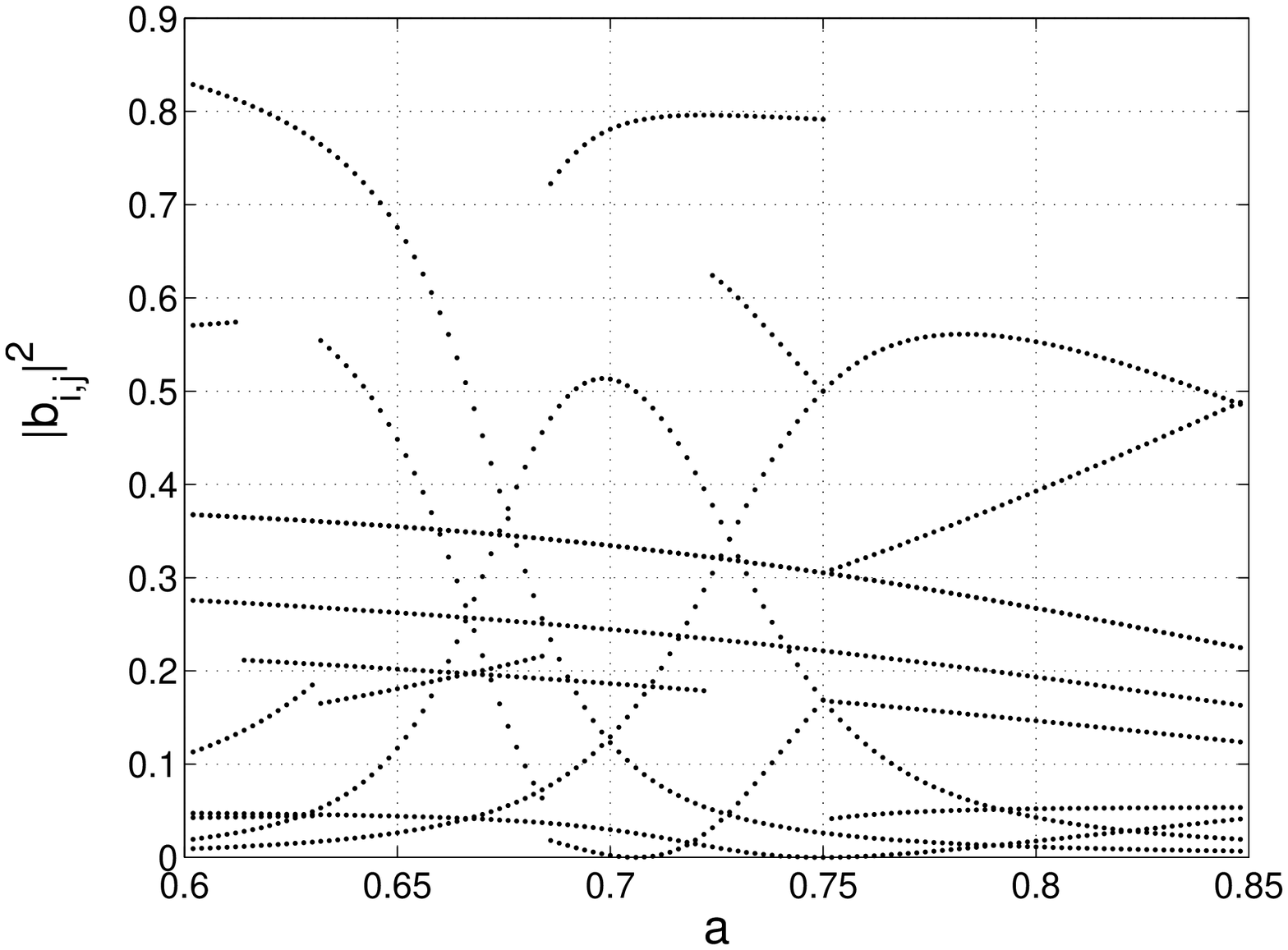,width=7.5cm}
\end{minipage}
\begin{minipage}[bl]{7.5cm}
\psfig{file=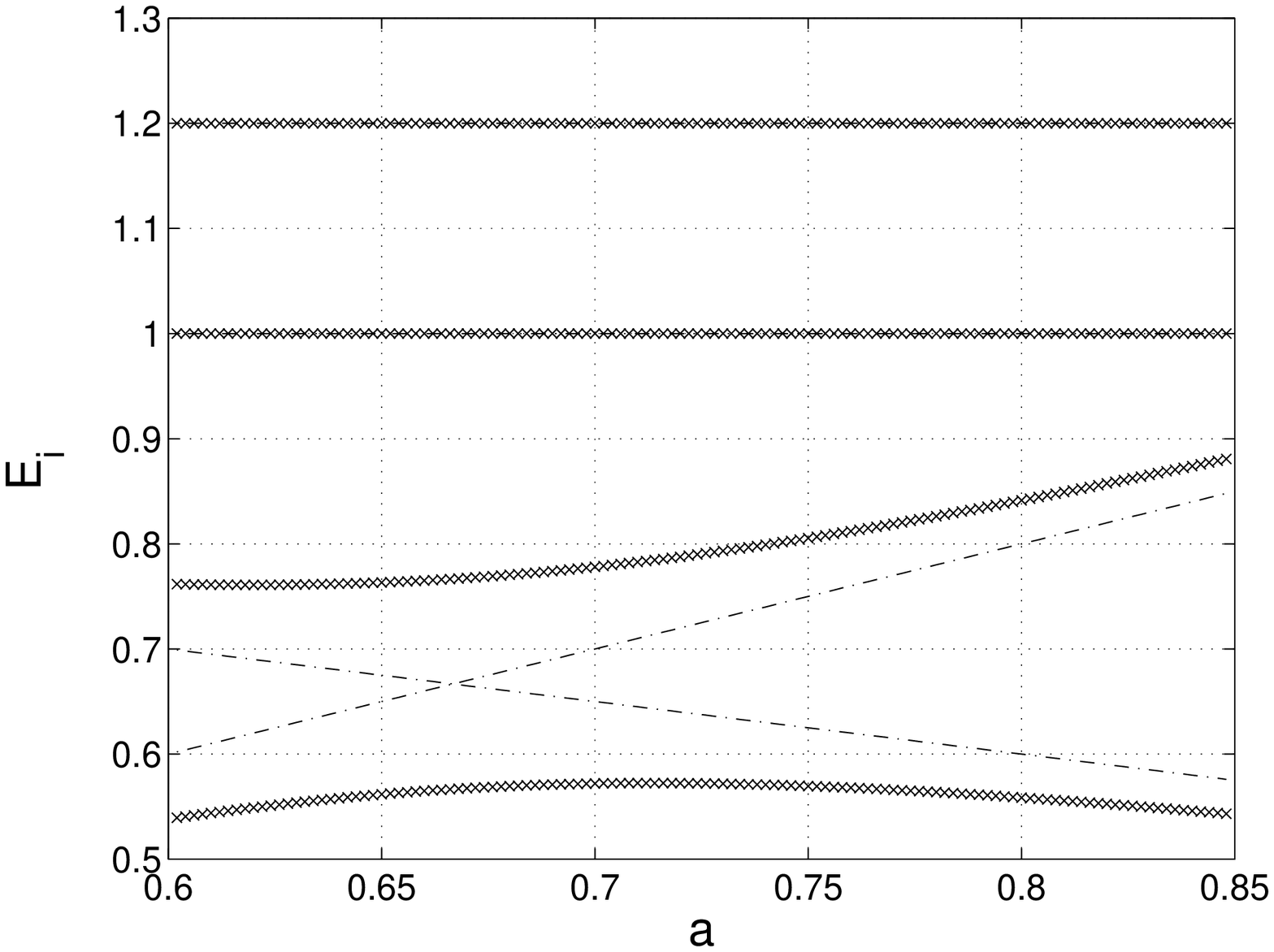,width=7.5cm}
\end{minipage}
\hspace*{1cm}
\begin{minipage}[br]{7.5cm}
\psfig{file=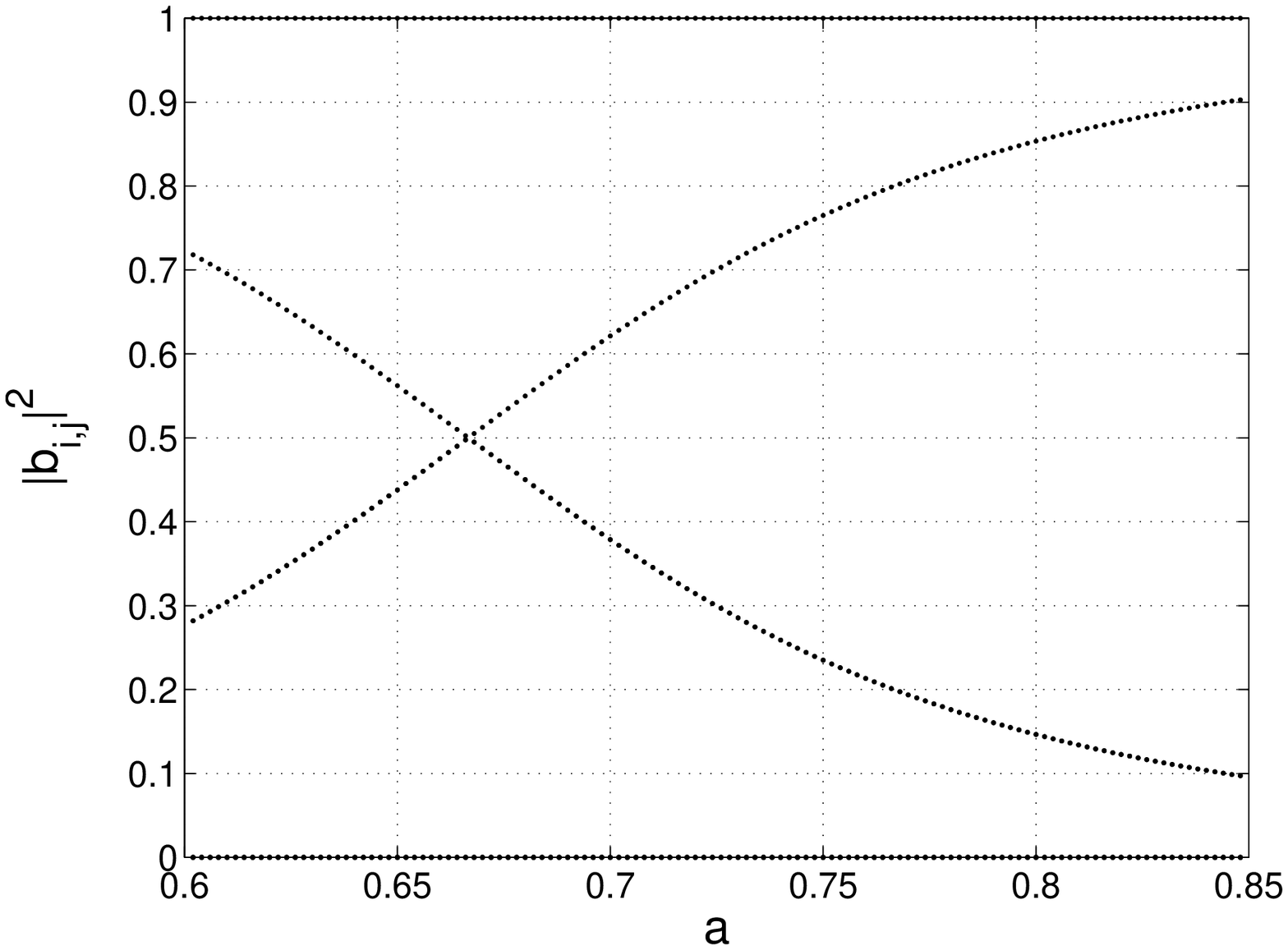,width=7.5cm}
\end{minipage}
\vspace*{.5cm}
\caption{The energies $E_i$ (left) and mixing coefficients $|b_{ij}|^2$
(right) of four discrete states ($\gamma_i = 0$ for $i= 1, ..., 4$)
obtained from ${\cal H}^{(4)}$, Eq. (\ref{eq:matr4}), as a function of the
tuning parameter $a$. Top and middle: $e_1=1-a/3; \; e_2=1- 5 a / 12 ; \; 
e_3=1-a/2; \;  e_4=a $; 
$~\omega = 0.05$ (top) and 0.1 (middle) for all non-diagonal matrix elements.
Bottom:  the same as above but 
 $e_1=1; \; e_2=1.2$;  $\omega =0$ for the coupling between
the states  $i=1,2$ and $j\ne i$,  $\; \omega =0.1$ for the coupling between
$i=3,4$ and $j=4,3$. In this case,  $|b_{ii}|^2  \ge |b_{ij\ne i}|^2$
(bottom right) as in Fig. \ref{fig:avoi2} (bottom right). 
The dash-dotted lines  (left)  show  $E_i$  for $\omega = 0$. 
The  states $i$ and $j$ are exchanged at some values $a$.
}
\label{fig:four}
\end{figure}

Another result of the present study is the influence of the branch 
points  in the complex plane  
onto the purity of the wave functions $\Phi_{i,j}$. 
At  $a^{\rm cr}$, the wave
functions are not only exchanged but become mixed. The mixing occurs not only
at the critical point  $a^{\rm cr}$ but in a certain region  around
$a^{\rm cr}$ when the crossing  is avoided. 
This fact is important at high level density where, as a rule, 
an avoided crossing with another level appears before $\Phi_i \to
\Phi^0_{j}$ is reached. As a result, all the wave functions of closely-lying
states contain components of all basic states. That means, they
 are strongly mixed at high level density 
(for illustration see Fig.  \ref{fig:four}). 

The strong mixing of the wave functions of a quantum system at high level
density means that the information on the individual properties of the  
discrete states described by the 
$\Phi_i^0$ is lost.  While the exchange of the wave
functions itself is of no interest for a statistical consideration of the
states, the accompanying mixing of the wave functions $\Phi_i$
is decisive for the 
statistics. At high level density, the number of branch points 
is relatively large (although of measure zero). 
Therefore, the (discrete  as well as resonance) states  
of quantum systems at high level density do not contain any
information on  the basic states with wave functions $\Phi_i^0$.  
It follows further that the statistical properties of quantum systems at 
high  level density are different from those at low level density. 
States at the border of the spectrum are almost not influenced by 
branch points in the complex plane since there are almost no states  which
could cross or  avoid crossing with others states. 
The properties of these states are
expected therefore to show more individual features than those at high level
density.  In other words, the information on the individual properties of the
states with wave functions $\Phi_i^0$ at the border of the spectrum  
is kept to a great deal in contrast to
that on the states in the center of the spectrum.

Thus, there is an influence of the continuum onto the properties
of a (closed) quantum system with discrete states
due to the analyticity of the wave functions. The branch 
points in the complex plane are {\it hidden crossings}, indeed. They play an
important role not only in atoms, as supposed in \cite{solov1,solov2}, 
but determine the 
properties of all (closed and open) quantum systems at high level density.
Their relation to  Berry phases has been studied experimentally as well as
theoretically  \cite{dubb,geom1,geom2}.
\\

Summarizing the results obtained for quantum systems
at high level density with avoided level
crossings and double poles of the $S$ matrix, it can be stated the following:
\begin{enumerate}
\item[--]
the poles of the $S$ matrix correspond to the eigenvalues of a non-hermitian 
effective Hamilton operator also in the case that the resonance
states overlap,

\item[--]
the  eigenfunctions of a non-Hermitian Hamilton operator 
are bi-orthogonal in the whole function space without any exceptions,

\item[--]
avoided level crossings in the complex plane as well as in the function
space of discrete states can be traced back  to the 
existence of branch points in the complex plane,

\item[--]
under certain conditions, a branch point in the complex plane appears as a
double pole of the $S$ matrix,

\item[--]
branch points in the complex plane  cause  an exchange of the wave
functions {\it and} create  a mixing of  the states of a 
quantum system at high level density even if the system is closed and the 
states are discrete.

\end{enumerate}

All the results obtained show the
strong influence of the branch points in the complex plane 
on the dynamics of many-level quantum systems.
They cause an avoided  overlapping of resonance states which
is accompanied by an exchange of the wave functions. A special case  is the
avoided level crossing of discrete states known for a long time.
The avoided level crossings cause a mixing of the 
eigenfunctions of ${\cal H}$ which is the larger the higher the level density
is. The states at the border of the spectrum of a many-particle system 
are therefore less influenced by avoided level crossings than those
in the centre.

\vspace*{.8cm}
\noindent
{\bf Acknowledgment}: Valuable discussions with 
E. Persson, J.M. Rost, 
 P. ${\rm \check S}$eba,  M. Sieber, E.A. Solov'ev
and H.J. St\"ockmann
are gratefully acknowledged.

\newpage

\end{document}